\documentclass[aps,prb,twocolumn,superscriptaddress,amsmath,amssymb,footinbib,longbibliography]{revtex4-2}
\usepackage{graphicx}   
\usepackage{verbatim}   
\usepackage{color}      
\usepackage{subfigure}  
\usepackage{hyperref}   
\usepackage{braket}		
\usepackage{mathrsfs}
\usepackage{ulem}
\newcommand{\op}[1]{\fontdimen12\textfont3=2pt\fontdimen12\scriptfont3=1.4pt\!\null\mathop{\protect\vphantom{#1}\smash{#1}}\limits_{\sim}\null\!}

\newcommand{\xref}[1]{\protect\ref{#1}}
\newcommand{\figref}[1]{Fig.~\protect\ref{#1}}
\newcommand{\fmref}[1]{(\protect\ref{#1})}

\renewcommand{\eqref}[1]{Eq.~(\protect\ref{#1})}

\def\erw#1{\langle \, {#1} \, \rangle}

\begin{document}

\title{Observation of phase synchronization and alignment during free induction decay of quantum spins with Heisenberg interactions}
\author{Patrick Vorndamme}
\affiliation{Faculty of Physics, Bielefeld University, 33615 Bielefeld, Germany}
\author{Heinz-J\"urgen Schmidt}
\affiliation{Department of Physics, Osnabr\"uck University, Barbarastra{\ss}e 7, 49076 Osnabr\"uck, Germany}
\author{Christian Schr\"oder}
\affiliation{Faculty of Physics, Bielefeld University, 33615 Bielefeld, Germany}
\affiliation{Bielefeld Institute for Applied Materials Research, Bielefeld University of Applied Sciences, 33619 Bielefeld, Germany}
\author{J\"urgen Schnack}
\email{jschnack@uni-bielefeld.de} 

\affiliation{Faculty of Physics, Bielefeld University, 33615 Bielefeld, Germany}

\date{\today}

\begin{abstract}
Equilibration of observables in closed quantum systems that are described by a unitary time evolution is a meanwhile 
well-established phenomenon apart from a few equally well-established exceptions. Here we report the surprising
theoretical observation that integrable as well as non-integrable spin rings with nearest-neighbor or long-range 
isotropic Heisenberg interaction not only
equilibrate but moreover also synchronize the directions of the expectation values of the individual spins. 
We highlight that 
this differs from spontaneous synchronization in quantum dissipative systems. 
Here, we observe mutual synchronization of local spin directions in closed systems
under unitary time evolution.
Contrary to dissipative systems, this synchronization is independent of 
whether the interaction is ferro- or antiferromagnetic.
In our numerical simulations, we investigate the free induction decay (FID) of an ensemble of up to 
$N = 25$ quantum spins with $s = 1/2$ each by solving the time-dependent Schr\"odinger equation numerically exactly.
Our findings are related to, but not fully explained by conservation laws of the system. The synchronization is 
very robust against for instance random fluctuations of the Heisenberg couplings and inhomogeneous magnetic fields.
Synchronization is not observed with strong enough 
symmetry-breaking interactions such as the dipolar interaction. 
We also compare our results to closed-system classical spin dynamics which 
does not exhibit phase synchronization due to the lack of 
entanglement. For classical spin systems the fixed magnitude of individual spins 
effectively acts like additional $N$ conservation laws.
\end{abstract}

\maketitle

\section{Introduction}
\label{sec-1}

Decoherence, equilibration as well as thermalization in closed quantum systems under unitary time evolution
are well-studied and by now well-established concepts which root in seminal papers by Deutsch, Srednicki and 
many others 
\cite{Deu:PRA91,Sre:PRE94,ScF:NPA96,RDO:N08,PSS:RMP11,ReK:NJP:12,SKN:PRL14,GoE:RPP16,AKP:AP16,BIS:PR16,WDL:PRB17,VoS:PRB20}.
For numerical studies, spin systems are the models of choice both since they are numerically feasible 
due to the finite size of their Hilbert spaces as well as they are experimentally accessible for
instance in standard investigations by means of electron parametric resonance (EPR), 
free induction decay (FID), or in atomic traps, see e.g.\
\cite{CAB:PRB97,MNR:PRB09,CMD:PRL09,ARM:PRL07,JHA:A21}.
In such systems, observables assume expectation values that are practically indistinguishable from 
the prediction of the diagonal ensemble for the vast majority of all times of their time evolution 
\cite{ReK:NJP:12,BaR:PRL17}.

In this paper we discuss an observation that rests both on decoherence and equilibration. 
We study the free induction decay (FID) of quantum spins that are arranged on a ring-like
geometry with nearest-neighbor as well as long-range isotropic Heisenberg interactions. 
For the overwhelming majority of investigated cases the initial product state of single-spin states
entangles, i.e. turns into a superposition of product states, and thereby equilibrates at the level of
single-spin observables. Our most striking observation is that expectation values of all individual
spin vectors synchronize with respect to their orientation. In a FID setting this means that their 
various individual rotations about the common field axis synchronize and align in the course of time.
In a co-rotating frame they simply align.
Experimentally, such collective effects may e.g.\ be imprinted in the temporal line shapes of the optical response 
under ultrashort pulse excitation and thus eventually be observed \cite{WMB:PRL92}.

We would like to contrast our findings with the longer-known observation of (spontaneous) synchronization 
in dissipative systems \cite{PhysRevB.82.144423,PhysRevA.88.042115,PhysRevLett.123.023604,PhysRevA.101.042121,Zhu_2015}. 
It was controversially discussed whether quantum two-level systems are able to synchronize at all \cite{PhysRevLett.121.053601},
with later conclusions that this is indeed the case \cite{PhysRevA.101.062104,PhysRevLett.123.023604}. 
All these investigations have in common that they try to identify stable limit cycles of the participating oscillators. 
Since such a discussion is applicable only to dissipative systems, which can emit or absorb energy to return to their stable
oscillation after a perturbation, it probably cannot serve as an explanation in our case.

A related and already investigated topic is transient synchronization in open quantum systems \cite{10.1007/978-3-030-31146-9_6}, in which the system finally equilibrates to a non-synchronized state,
but synchronizes temporarily on the way. We show that we observe a comparable behavior in closed quantum spin systems,
if we weakly reduce the symmetry of the Hamiltonian.

The observed synchronization is stable against random fluctuations of the Heisenberg couplings
and we observe it for almost all initial conditions.
We therefore conjecture that it is tightly connected to the symmetries and conserved quantities 
of the isotropic Heisenberg model which is SU(2) invariant \cite{ScS:IRPC10}, 
see also \cite{UHS:PRB14}. 
This hypothesis is corroborated by the observation that strongly anisotropic interactions
such as the dipolar interaction spoil the synchronization. Also in classical spin dynamics the phenomenon 
cannot be observed as will be discussed in detail later. 
Inhomogeneous or randomly fluctuating local fields at the sites of the
individual spins on the other hand do not prevent the spins from synchronizing although the 
conservation laws are broken. The same applies for weakly anisotropic interactions that are close to the 
isotropic Heisenberg case. We observe a transient synchronization.

The paper is organized as follows. In Sect.~\ref{sec-2} we
introduce the theoretical model and the applied methods. 
Section~\ref{sec-3} deals with exemplary numerical quantum simulations under isotropic Heisenberg interactions 
and we compare to classical simulations. Section~\ref{sec-4} introduces symmetry breaking anisotropic interactions 
and demonstrates the transient behaviour of the synchronization phenomenon.
Section~\ref{sec-5} provides a summary of our main results.
In the appendix some aspects are discussed in more detail, especially the behavior under symmetry breaking interactions.
Video clips of our simulations are provided on the website of the paper \cite{VSS:21}.

\section{Theoretical model and methods}
\label{sec-2}

The Hamiltonian of our spin model reads
\begin{align}
\op H =  - \sum_{j = 1}^N  J_j \vec{\op {s}}_j \cdot \vec{\op s}_{j+1}  - \sum_{j = 1}^N h_j \op s_j^z
\ ,
\label{H1}
\end{align}
where the first sum corresponds to the isotropic Heisenberg model and the second sum denotes the Zeeman term.
Operators are marked by a tilde, the Heisenberg interactions are denoted by $J_j$, local magnetic fields are given by $h_j$,
and periodic boundary conditions $\vec{\op {s}}_{N+1} = \vec{\op {s}}_{1}$ are applied. 
Thus, the Hamiltonian describes spins which are 
arranged as a ring; it could for instance be a ring molecule \cite{TGA:PRL05,UNM:PRB12}. 
We define the total spin operator
\begin{align}
\vec{\op S} := \sum_{k = 1}^N  \vec{\op s}_k
\ ,
\label{eqGesamtspin}
\end{align}
which commutes with the Heisenberg part of the Hamiltonian, and so does $\vec{\op S^2}$, 
even if the coupling constants $J_j$ are all different. This is true for any spin arrangement, 
not just for rings \cite{BSS:JMMM00,ScS:IRPC10}.
The conservation of $\vec{\op S}^2$ is broken either by anisotropic 
interactions or by varying local magnetic fields $h_j$
\begin{align}
\left[\vec{\op S}^2, \sum_{i = 1}^N  h_i \op s_i^z \right] \propto \underset{= \; 0 \; \textsf{only if} \;  h_i = h_j}
{\left[\vec{\op s}_i \cdot \vec{\op s}_j, \; \; h_i \op s_i^z + h_j  \op s_j^z \right]}
\ .
\label{eqkommu2}
\end{align}

Furthermore, we define the transverse magnetization
\begin{align}
M_{\text{trans}} :&= \sqrt{ \erw{\op S^x}^2 + \erw{\op S^y}^2 } \notag \\
			&= \sqrt{ \Bigg(\sum_j \erw{\op s_j^x} \Bigg)^2 + \Bigg(\sum_j \erw{\op s_j^y} \Bigg)^2 }
			\ . 
\label{Mtrans}
\end{align}
Here $\erw{\op S^x}$ denotes the expectation value with respect to a specified many-body state.
We interpret \fmref{Mtrans} as the net magnetization precessing in the $xy$-plane. 
In case of Hamiltonian \fmref{H1} this is also a conserved quantity 
if the local magnetic fields are all the same $h_j \equiv h \; \forall j$. 
This can be seen by looking at the time evolution ($\hbar := 1$)
\begin{align}
 \frac{d}{dt} \bra{\psi(t)}   \vec{ \op S}  \ket{\psi(t)} &= \frac{1}{\mathrm{i} } \bra{\psi(t)} [ \vec{\op S}, \op H]  \ket{\psi(t)} \notag \\
		&= \mathrm{i}  h \bra{\psi(t)} [ \vec{\op S}, \op S^z]  \ket{\psi(t)}
		\ .
\label{Mtranst}
\end{align}
Remember, $h$ denotes the magnetic field. 
The solution of \eqref{Mtranst} is of the form 
\begin{align}
 \bra{\psi(t)}   \vec{ \op S}  \ket{\psi(t)} = \left(\begin{array}{c} 
      										a \cos{h t} + b \sin{h t}\\  -b \cos{h t} + a \sin{h t} \\ c 
   										 \end{array}\right)
   										 \ ,
\label{Mtranst2}
\end{align}
compatible with the conserved quantities.
The coefficients $a$, $b$ and $c$ are determined by the initial state of the system. 
We observe a collective 
rotation with frequency $h$ in all cases the spins synchronize (Sections
\ref{sec-3-1}, \ref{sec-3-2} and \ref{sec-3-4}). 
Appendix \ref{app-2-1} provides an exception where the spins collectively 
precess around a mean field $\tilde{h}$. 

\begin{figure}[h!]
\centering
\includegraphics*[clip,width=0.95\columnwidth]{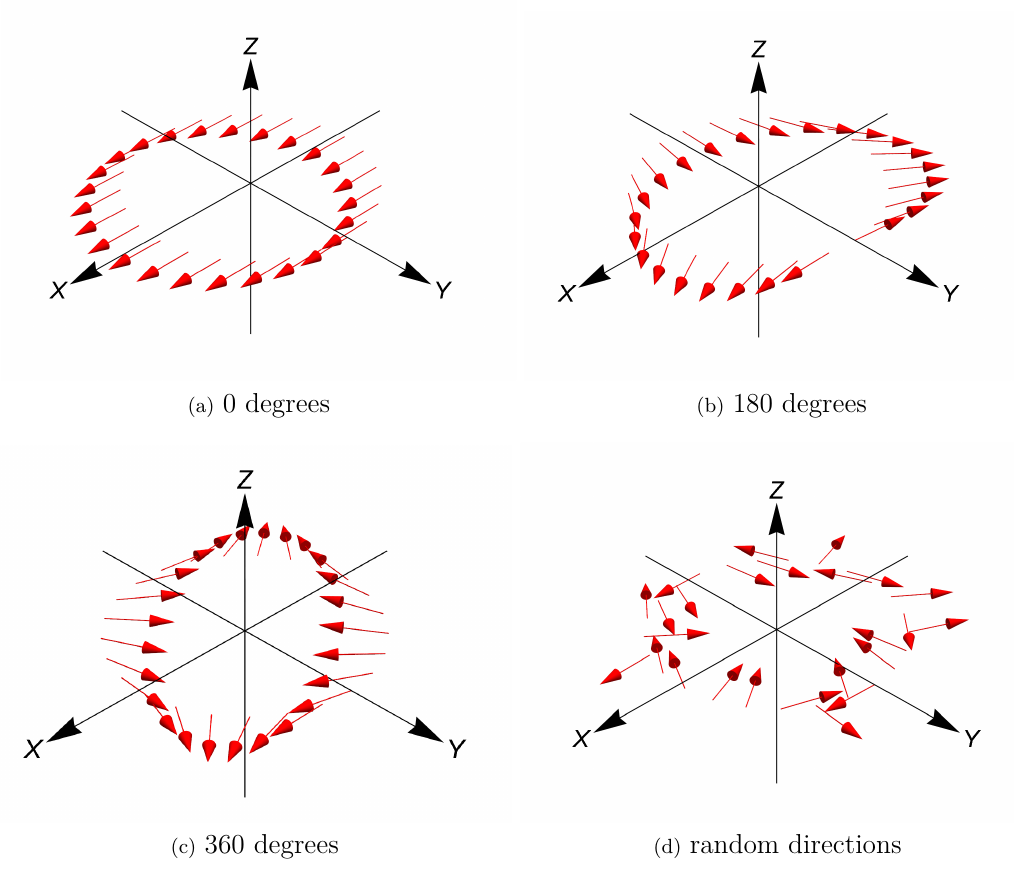}
\caption{\label{fig:1} Visualization of the four initial product states
studied in this work. The arrows correspond to the single-spin expectation values,
see text and \eqref{psi0}. We will refer to this states as $\ket{\psi_A}$, $\ket{\psi_B}$, 
$\ket{\psi_C}$, and $\ket{\psi_D}$.}
\end{figure}

As initial many-body states we choose product states of the form
\begin{equation}
 \ket{\psi(t=0)} = \bigotimes\limits_{j=1}^{N} \frac{1}{\sqrt{2}} \left(\ket{\uparrow} + e^{i \theta_j} \ket{\downarrow}  \right) 
\ ,
\label{psi0}
\end{equation}
for which the expectation values of individual spins
\begin{align}
\erw{\vec{\op s}_j} := \bra{\psi} \vec{\op s}_j \ket{\psi}
\label{spinExpectation}
\end{align}
are oriented in the $xy$-plane and point in a direction that depends on $\theta_j$.
In the following we are going to investigate the time evolution of the four 
states shown in \figref{fig:1} where (a) all spins point in the same direction, 
(b) are regularly fanned out by $180$ degrees, 
(c) are regularly fanned out by $360$ degrees, and (d) point in random directions.
We will refer to this states as $\ket{\psi_A}$, $\ket{\psi_B}$, 
$\ket{\psi_C}$, and $\ket{\psi_D}$ (or A, B, C, and D in the classical case, sec.~\xref{sec-3-5}).

In a product state, the spins are not entangled by definition, however they entangle 
during the unitary time evolution
\begin{equation}
\ket{\psi (t)} = e^{-\mathrm{i} \op H t} \ket{\psi (0)}
\ ,
\label{schrödi2}
\end{equation}
that we calculate numerically exactly using a Suzuki Trotter decomposition \cite{Hatano2005}. 

In order to measure the entanglement of an individual spin at site $j$ with the others, we 
define the reduced density matrix
\begin{equation}
\op \rho_j = \textsl{Tr}_{\bigotimes\limits_{k \neq j} \mathscr{H}_k} \left( \ket{\psi} \bra{\psi} \right)
\ .
\label{density}
\end{equation}

Here $\mathscr{H}_j$ denotes the Hilbert subspace of spin $j$, and 
$\mathscr{H} = \bigotimes\limits_{j=1}^{N} \mathscr{H}_j$ is the total Hilbert space.
The purity is given as $\textsl{Tr}   \left( \op \rho_j^2  \right)$. 
$\textsl{Tr}   \left( \op \rho_j^2  \right) = 1$ holds, if spin $j$ is 
not entangled with other spins, and  $\textsl{Tr}   \left(\op \rho_j^2  \right) = 0.5$ 
if it is maximally entangled with other spins. 
The purity is thus also a measure of decoherence for an observer of a single spin 
\cite{RevModPhys.76.1267,2014arXiv1404.2635S}. An alternative way of quantifying the decoherence 
would be the von Neumann entropy $S( \op \rho_j) = - \textsl{Tr} \left( \op \rho_j \log_2 \op
\rho_j \right)$ \cite{SciPostPhys.2.2.010}.

\section{Calculations and results}
\label{sec-3}

In this Section we present our numerical findings of the special behaviour of initial states in 
\figref{fig:1} under time evolution with Hamiltonian \eqref{H1} and equal magnetic fields $h_j = -1 \;\forall j$.
As discussed, $M_{\text{trans}}$ and $\vec{\op S^2}$ are conserved quantities. 
We show that, with one exception, the spin expectation
values synchronize.

\subsection{Initial state $\ket{\psi_A}$}
\label{sec-3-1}

\begin{figure}[h!]
\centering
\includegraphics*[clip,width=1.0\columnwidth]{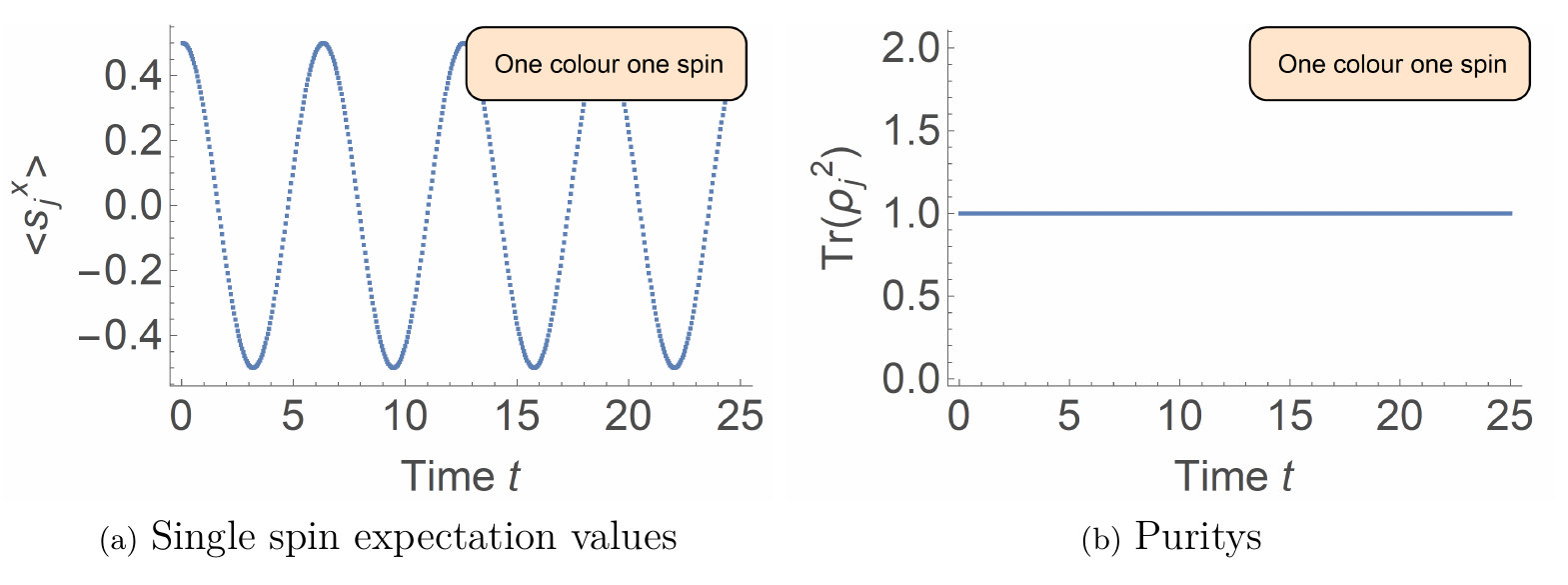}
\caption{\label{fig:2} Time evolution of initial state $\ket{\psi_A}$ regarding Hamiltonian \eqref{H1} 
with isotropic Heisenberg interactions 
and $J_j \in {[1.6,2.4]}$, $h_j = -1 \;\forall j$, $N = 25$. Left panel: Single-spin expectation values in $x$-direction. 
Right panel: Purity of the individual reduced density matrices.}
\end{figure}

In \figref{fig:2} we start with initial state $\ket{\psi_A}$ and random Heisenberg interactions $J_j$. 
In this case, every spin is precessing as if independent 
without entangling to other spins, no matter how the $J_j$ are chosen. 
Since all spins point in 
the same direction, $M_{\text{trans}}$ and $\vec{\op S^2}$ assume
their maximum values. Because they are conserved quantities the spins are bound to 
remain in a perfect product state, otherwise it would not be possible 
to conserve these values over time.

\subsection{Initial state $\ket{\psi_B}$}
\label{sec-3-2}

\begin{figure}[h!]
\centering
\includegraphics*[clip,width=1.0\columnwidth]{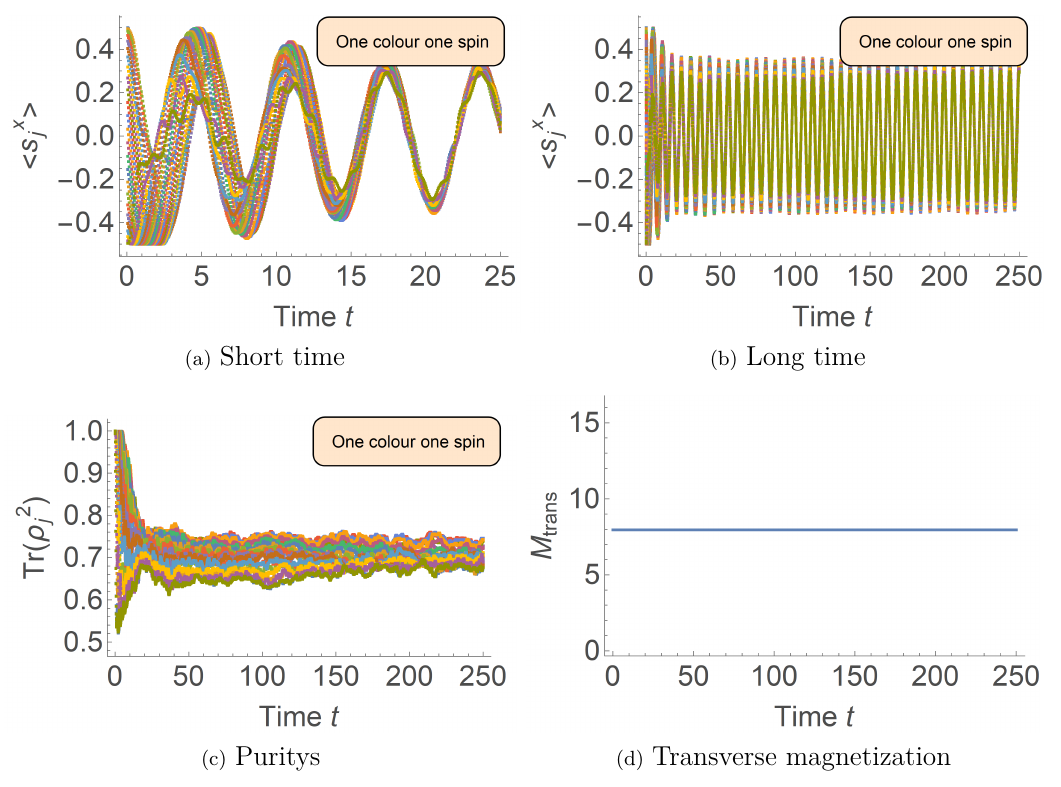}
\caption{\label{fig:3} Time evolution of initial state $\ket{\psi_B}$ 
w.r.t.\ Hamiltonian \eqref{H1} with 
isotropic Heisenberg interactions 
and $J_j \in {[1.6,2.4]}$, $h_j = -1 \;\forall j$, $N = 25$.
The video for \xref{fig:3}(a) can be found at \cite{VSS:21}.}
\end{figure}

Figure~\xref{fig:3} shows almost the same as \figref{fig:2}, but this time for initial state 
$\ket{\psi_B}$. Initially the individual spin expectation values 
are spread out by $180$ degrees, but during time evolution they align.
This astonishing phenomenon can be nicely observed in the video provided 
on the web page of the published article \cite{VSS:21}.

During time evolution and synchronization the spins entangle as much as the conservation 
of $\vec{\op S}^2$ and  $M_{\text{trans}}$ allows. Interestingly, 
the spins stay entangled and do not fan out again
(apart from finite size effects such as revivals at very late times). 
This statement becomes stronger with increasing system size,
which is further addressed in Appendix \ref{app1}. 
We interpret this phenomenon as quantum mechanical equlibration process
under the restricting influence of conserved quantities \cite{UHS:PRB14}.

The synchronization can be rationalized for spin systems where all spins are equivalent, 
i.e.\ ring systems with translational invariance ($J_j=J$, $h_j=h\;\forall j$) since then
equilibration should result in the same single-spin expectation value at every site.
This concerns magnitude and direction of the spin vector. The somewhat unexpected result
of our investigation is that the direction of all spins continues to synchronize also
for settings where spins are no longer equivalent, i.e.\ if the Heisenberg interactions are drawn 
at random from a distribution.

Figure~\xref{fig:3}(c) shows the purity of the individual reduced density operators $\op \rho_j$ (\eqref{density}). 
Since the couplings $J_j$ are different for different $j$, not all spins are equal. This does not prevent the spins
from synchronizing their directions, but they do not all entangle to the same extent.

Another main result of this paper is that the time
needed for the spins to synchronize is almost independent 
of the width $\Delta$ of the distribution of the $J_j  \in {[2 - \Delta,2 + \Delta]}$.
This is also demonstrated numerically in Appendix \ref{app1}.

\begin{figure}[h!]
\centering
\includegraphics*[clip,width=1\columnwidth]{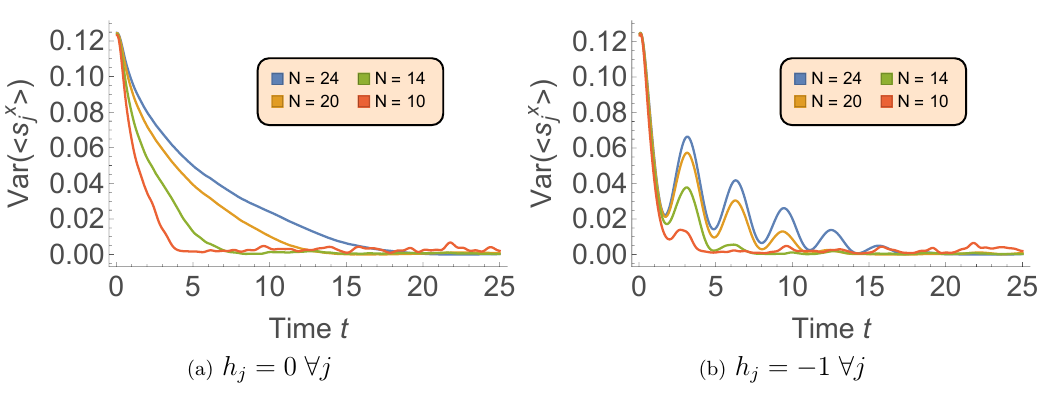}
\caption{\label{fig:4} Time evolution of initial state $\ket{\psi_B}$ 
w.r.t. Hamiltonian \eqref{H1} 
with isotropic Heisenberg interactions 
and $J_j = 2 \;\forall j$ without (a) and with magnetic field (b).}
\end{figure}

Figure~\xref{fig:4} shows the variance of the expectation values of individual spin operators, defined as
\begin{align}
\textsf{Var}(<\op s_j^x>)(t) := 
\frac{1}{N}
\sum_{j=1}^N \left(<\op s_j^x> - \frac{<S^x>}{N}\right)^2
\label{eqVar}
\end{align}

for different system sizes $N$. 
That the variance decays to zero, compare \figref{fig:4}, 
expresses precisely that the spins align 
until they point in the same direction. 
This process takes the longer the larger the system is.
The synchronisation, i.e.\ the alignment of directions, 
also takes place in the absence of a magnetic field, as can be seen 
in \figref{fig:4}(a). 
The reason is that the homogeneous magnetic field, which is a one-body operator, 
does not causes any many-body entanglement between the spins; entanglement and 
equilibration are driven by the Heisenberg term which is a two-body operator.
As a result, the field-free curves in \figref{fig:4}(a) are the envelopes of the curves
taken with homogeneous field and shown in \figref{fig:4}(b).

\subsection{Initial state $\ket{\psi_C}$}
\label{sec-3-3}

\begin{figure}[h!]
\centering
\includegraphics*[clip,width=1\columnwidth]{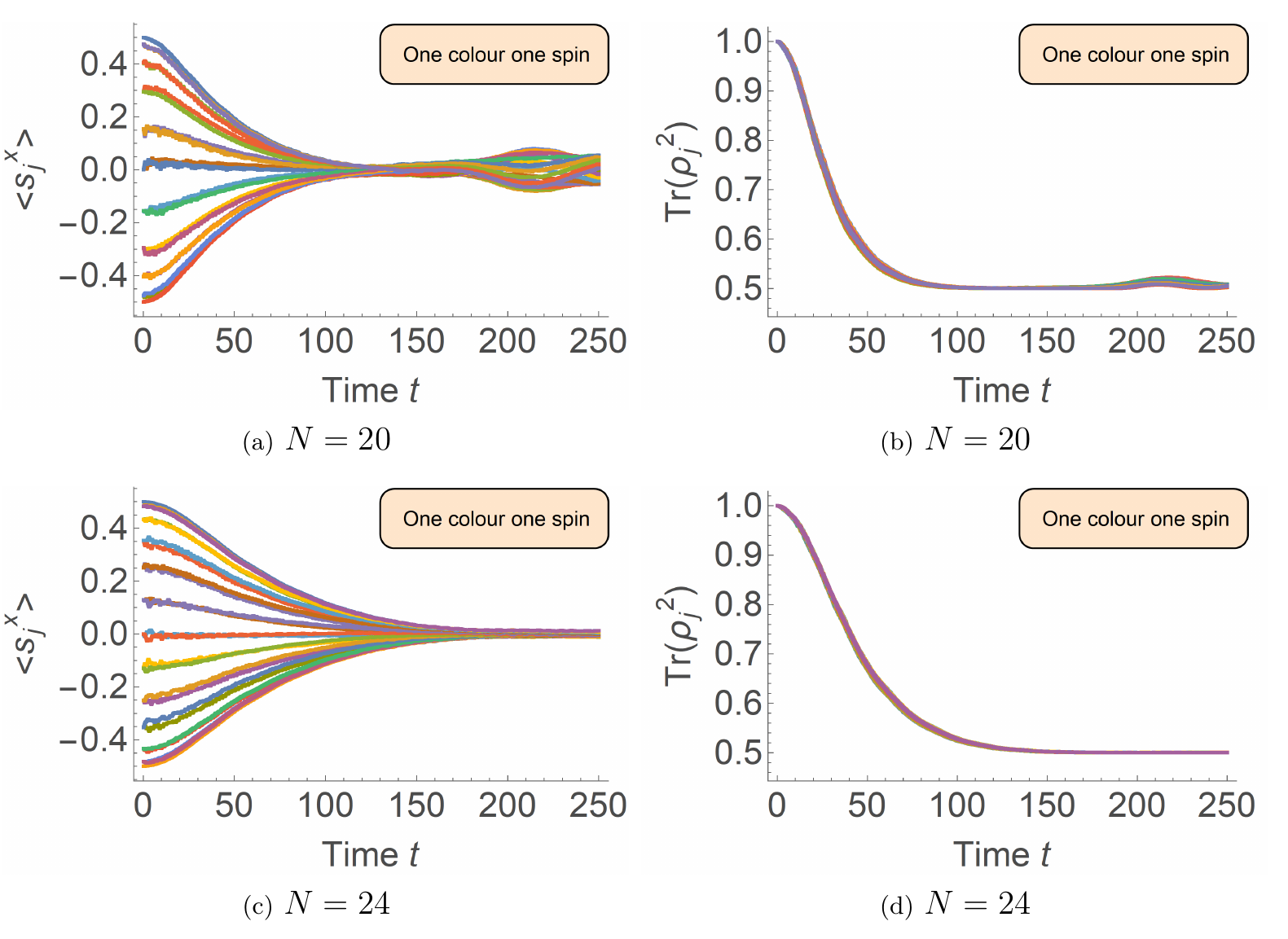}
\caption{\label{fig:5}Time evolution of initial state $\ket{\psi_C}$ w.r.t.\ Hamiltonian 
\eqref{H1} with isotropic Heisenberg interactions and $J_j  \in {[1.6,2.4]}$, $h_j = 0 \;\forall j$
for different system sizes. The video of \xref{fig:5}(c) is provided at \cite{VSS:21}.}
\end{figure}

Figure~\xref{fig:5} shows the time evolution for initial state $\ket{\psi_C}$ and different system 
sizes. This is a very special and atypical case with a particular symmetry in the spin orientations  
which  results in a very stable state even if there are different couplings $J_j$ between the spins,
see video \cite{VSS:21}. This is the only initial state we find where the spins do not align, but 
entangle and decay to zero, with wild echos at later times. The larger the system, the longer 
the echos take to occur and the longer it takes for the spins to entangle.  

These numerical results for finite system sizes suggest that $\ket{\psi_C}$ is an energy eigenstate in the 
thermodynamic limit which appears plausible, because the angle between neighboring spins is given by $2 \pi/N$,
therefore for $N \rightarrow \infty$ all neighbors are parallel in the initial state. 
We emphasize that this state would also be an energy eigenstate in the non-integrable case \cite{Bab:NPB83,FYF:NPB90}
where the $J_j$ are all different. 
Because all single-spin observables are strongly different we conjecture that this state is not thermal; 
its relation to quantum scars needs to be explored,
see \cite{CTP:PRL19,PVB:NC21} and references therein.

\subsection{Initial state $\ket{\psi_D}$}
\label{sec-3-4}

\begin{figure}[h!]
\centering
\includegraphics*[clip,width=1\columnwidth]{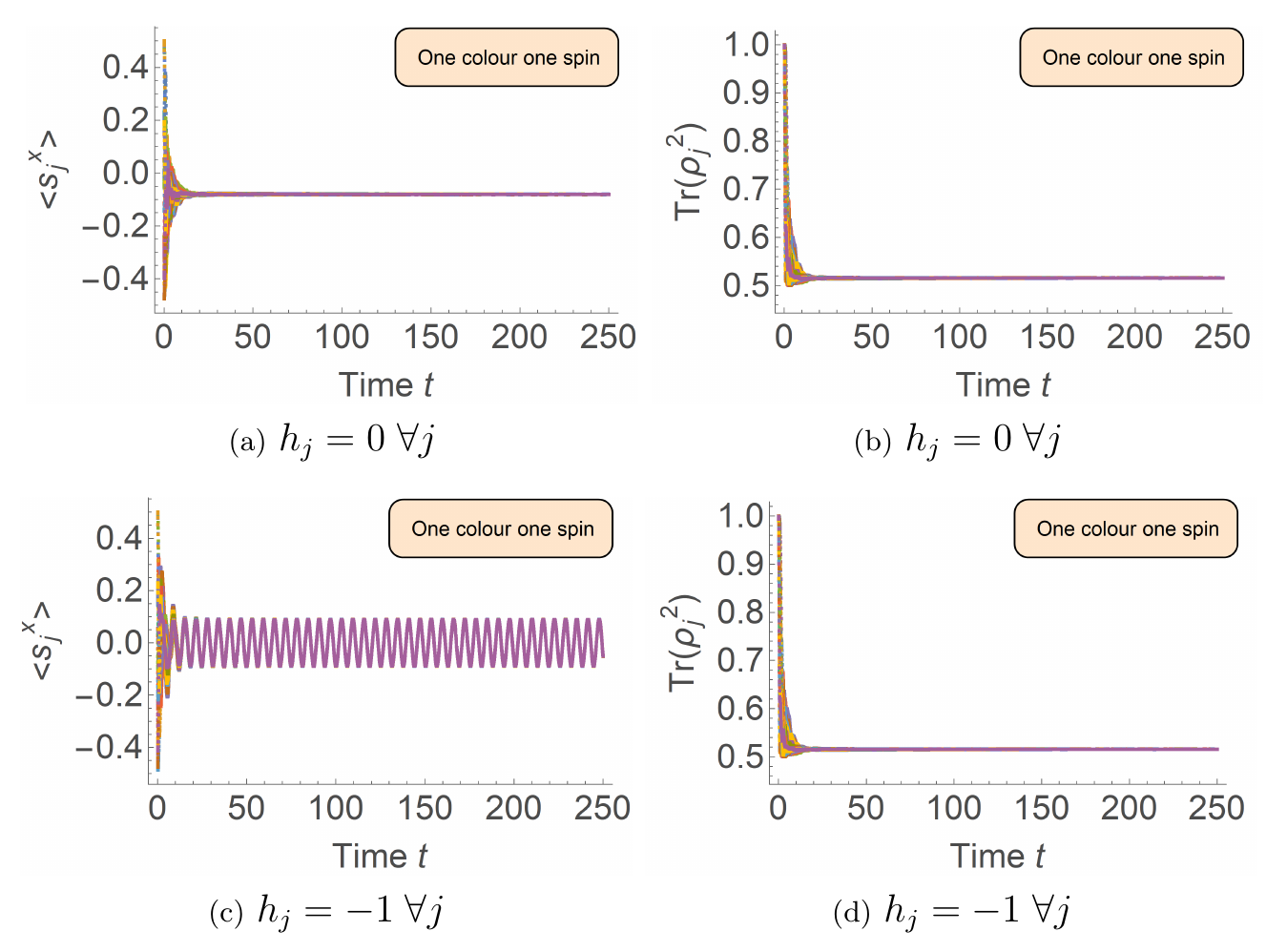}
\caption{\label{fig:6}Time evolution of initial state $\ket{\psi_D}$ w.r.t.\
Hamiltonian 
\eqref{H1} with isotropic Heisenberg interactions and $J_j \in {[1.6,2.4]}$, $N = 24$
without (a-b) and with magnetic field (c-d).
Video \xref{fig:6}(c) is provided at \cite{VSS:21}.}
\end{figure}

Figure \xref{fig:6} shows a time evolution for initial state $\ket{\psi_D}$ (random orientations) 
without magnetic field (\figref{fig:6}(a)) and with magnetic field (\figref{fig:6}(b)). The 
conserved net magnetization is small (would be zero in the thermodynamic limit or as a mean of many 
random realizations according to the central limit theorem).
Nevertheless, the spins synchronize which shows that this phenomenon is very 
robust with respect to the initial state, see also video \cite{VSS:21}.

\subsection{Classical spin dynamics}
\label{sec-3-5}

\begin{figure}[ht!]
\centering
\includegraphics*[clip,width=1.0\columnwidth]{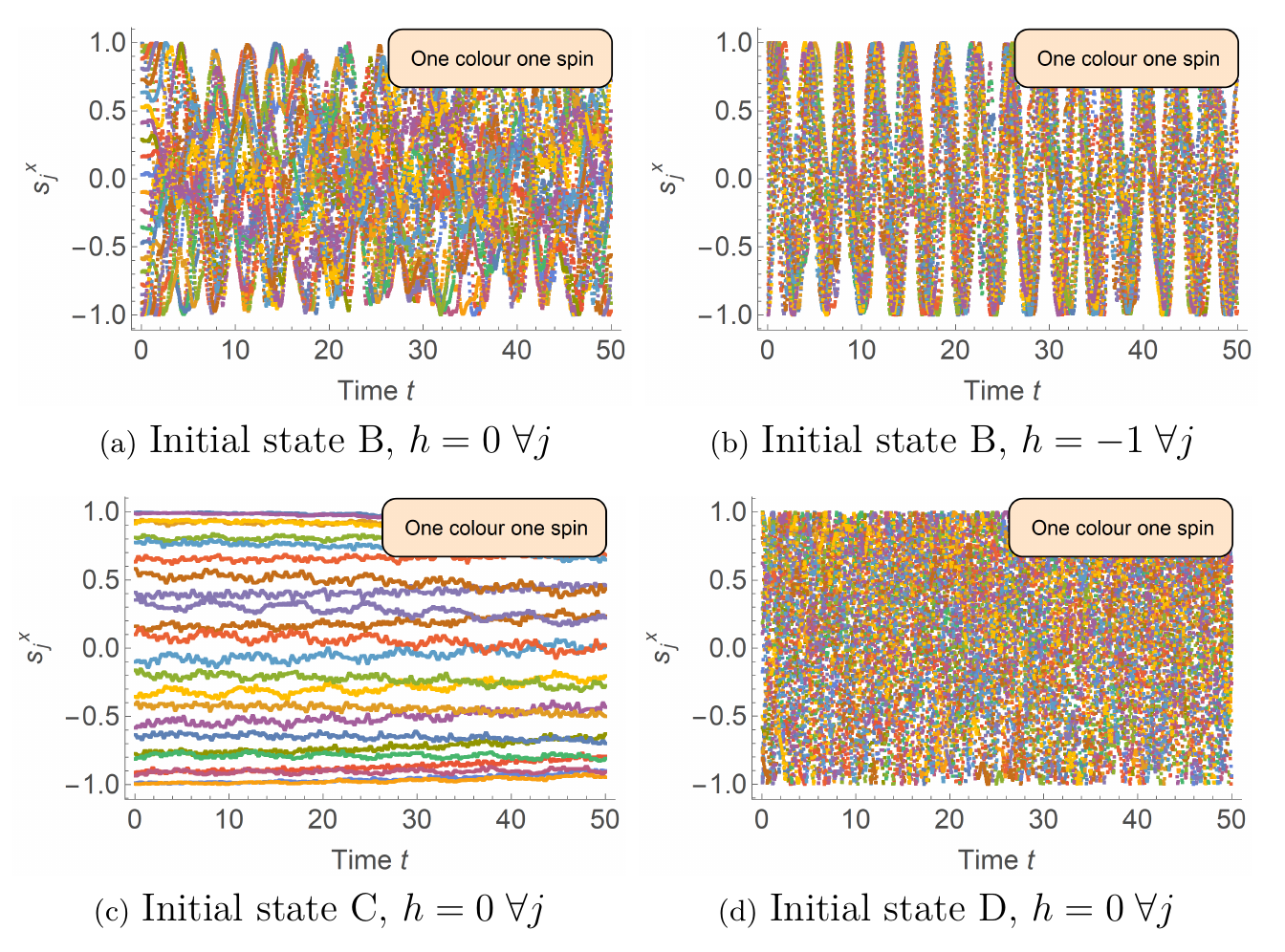}
\caption{\label{fig:7}Classical time evolution for various initial states w.r.t.\
Hamiltonian 
\eqref{eqKlassisch} with $N = 24$, $J_j \in {[1.6,2.4]}$.
Videos for all cases are provided at \cite{VSS:21}.}
\end{figure}

Although not at the heart of our investigations, we like to
compare our results to classical spin dynamics with identical Heisenberg couplings and initial states.
It turns out that a classical spin dynamics does not exhibit phase synchronization,
see videos \cite{VSS:21}. 
The reason in this context is that classical spin dynamics lacks entanglement 
which is necessary for synchronization.
Contrary to the expectation values of the quantum spins, the classical spins are bound
to maintain their length, which effectively acts like additional $N$ conservation laws. 
This results in a oscillatory dynamics of most investigated initial conditions, compare 
\figref{fig:7} for initial states equivalent to \figref{fig:2}(b)-(d). 
Initial state C is again special. In the quantum case the spins maintain 
their directions for a long time while slowly entangling. In the classical case the spins 
also keep their directions for a long time. 
This is because of the high symmetry in the initial state with zero net magnetization 
and no preferred direction in the $x$y-plane.

The classical spin dynamics has been evaluated according to
\begin{align}
\frac{d}{dt} \vec{s}_j
=
\frac{\partial H}{\partial \vec{s}_j} \times \vec{s}_j
\ ,
\label{eqKlassisch}
\end{align}
where $H$ denotes the classical Hamiltonian analogous to \eqref{H1}.

In the literature one finds well-studied classical examples of synchronization of e.g.\ Van der Pol oscillators \cite{doi:10.1080/14786442608564127}. These are dissipative systems with stable limit cycles, which means that
after a perturbation they return to their ordinary oscillation. 
This would not be the case for our classical spins even if we would couple them to a heat bath.

One trivial (but much different) way of how our classical spins would synchronize (at least in a transient way) 
is by choosing a ferromagnetic Heisenberg coupling $J_j$ and a dissipative dynamics so that the ferromagnetic ground state 
is approached which is aligned trivially.

\begin{figure}[ht!]
\centering
\includegraphics*[clip,width=1.0\columnwidth]{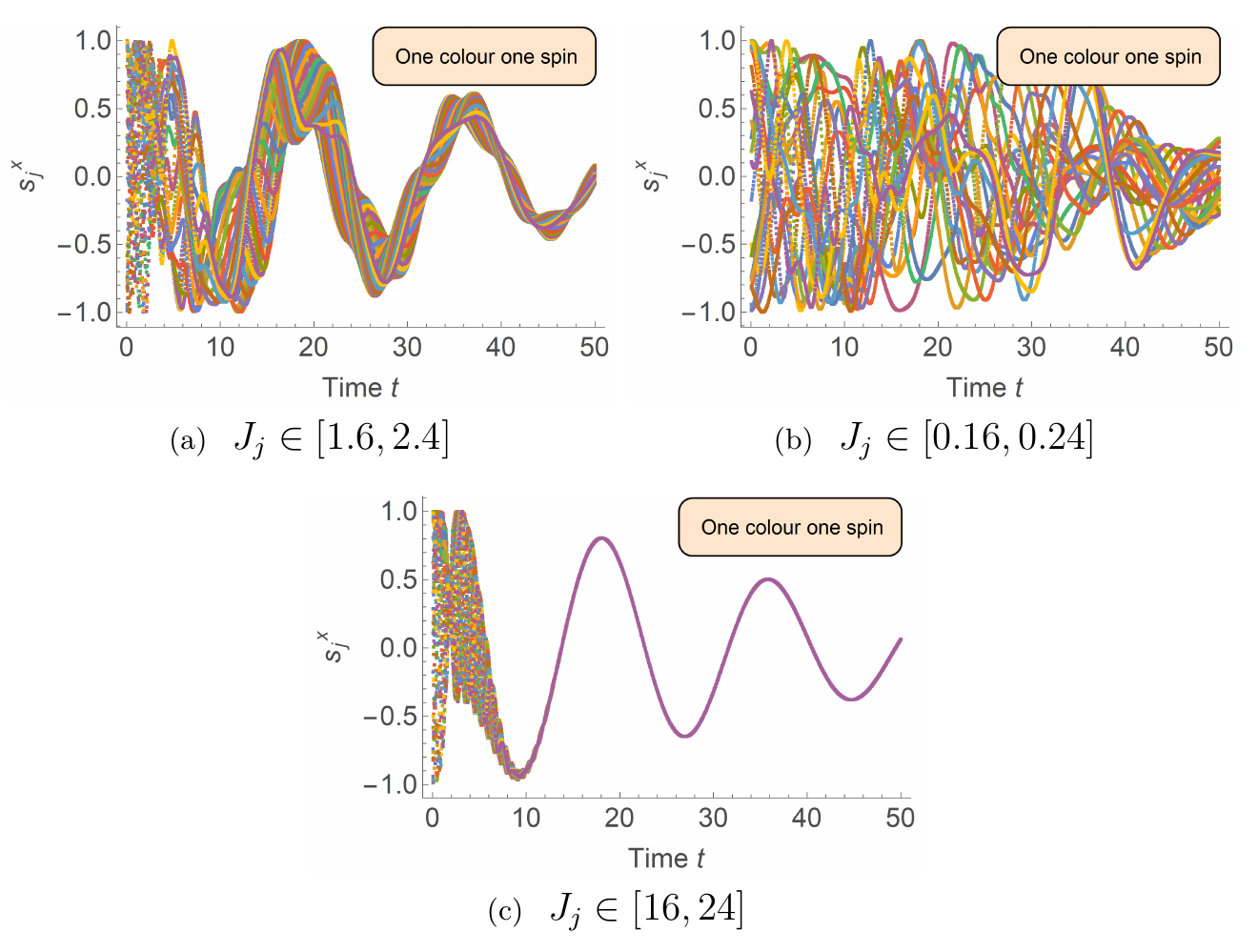}
\caption{\label{fig:8} Damped classical time evolution for initial state B w.r.t.\
Hamiltonian 
\eqref{eqKlassisch2}, $\alpha = 0.1$, $N = 24$, $h_j = -1 \;\forall j$ and different intervals of the ferromagnetic coupling strength.}
\end{figure}

Figure~\xref{fig:8} shows such an example for different intervals of the Heisenberg coupling strength. 
The damping is realized by
\begin{align}
\frac{d}{dt} \vec{s}_j
=
\frac{\partial H}{\partial \vec{s}_j} \times \vec{s}_j - \alpha \left( \frac{\partial H}{\partial \vec{s}_j} \times \vec{s}_j \right) \times \vec{s}_j
\ .
\label{eqKlassisch2}
\end{align}

Because of the damping, the system looses energy until it arrives in the lowest possible energy state. 
Since a ferromagnetic coupling is chosen, the spins synchronize per default. 
In \figref{fig:8} there is also a magnetic field applied in $z$-direction 
with which the spins also align. Therefore, at late times all spins point in $z$-direction. 
If the ferromagnetic coupling is much larger than the magnetic field $|J_j| \gg |h|$, 
then also the $x$- and $y$-components of the spins synchronize in a transient way.

We note that such kind of synchronization is fundamentally different from the synchronization 
we observe in closed quantum spin systems. In the quantum case, it does not matter 
if the Heisenberg interaction is ferro- or antiferromagnetic. 
The synchronization is just based on equilibration, entanglement, and conservation laws.

\section{Breaking the symmetry}
\label{sec-4}

In this Section we investigate whether synchronization still occurs 
if $M_{\text{trans}}$ and $\vec{\op S}^2$ are not conserved anymore. 
We can break the symmetry in different ways, 
either by means of inhomogeneous magnetic fields 
or by interactions between the spins that are not of isotropic Heisenberg type. 
In Section (\ref{sec-4-1}) we choose XYZ interactions and in 
Section (\ref{sec-4-2})  XX interactions as two examples with different outcomes. 
In Appendix \ref{app2} we show the effect of 
inhomogeneous magnetic fields (App. \ref{app-2-1}) and 
of dipolar interactions between all spins (App. \ref{app-2-2}).

\subsection{XYZ interaction}
\label{sec-4-1}

We begin with the XYZ interaction which is close to the isotropic Heisenberg 
case if the interaction 
in the three spatial directions is not too different. In this
case, the synchronization between the spins still occurs. 
The Hamiltonian in this subsection is defined as
\begin{align}
\op H_{XYZ} =  &- J \sum_{j = 1}^N \op s_j^x \op s_{j+1}^x - (J - \delta) \sum_{j = 1}^N \op s_j^y \op s_{j+1}^y  
\notag \\ 
&- (J - 2\delta) \sum_{j = 1}^N \op s_j^z \op s_{j+1}^z - h \sum_{j = 1}^N  \op s_j^z \; .
\label{Hxyz}
\end{align}

We use the parameter $\delta$ to tune the difference of the interaction in the three spatial directions.

\begin{figure}[h!]
\centering
\includegraphics*[clip,width=1\columnwidth]{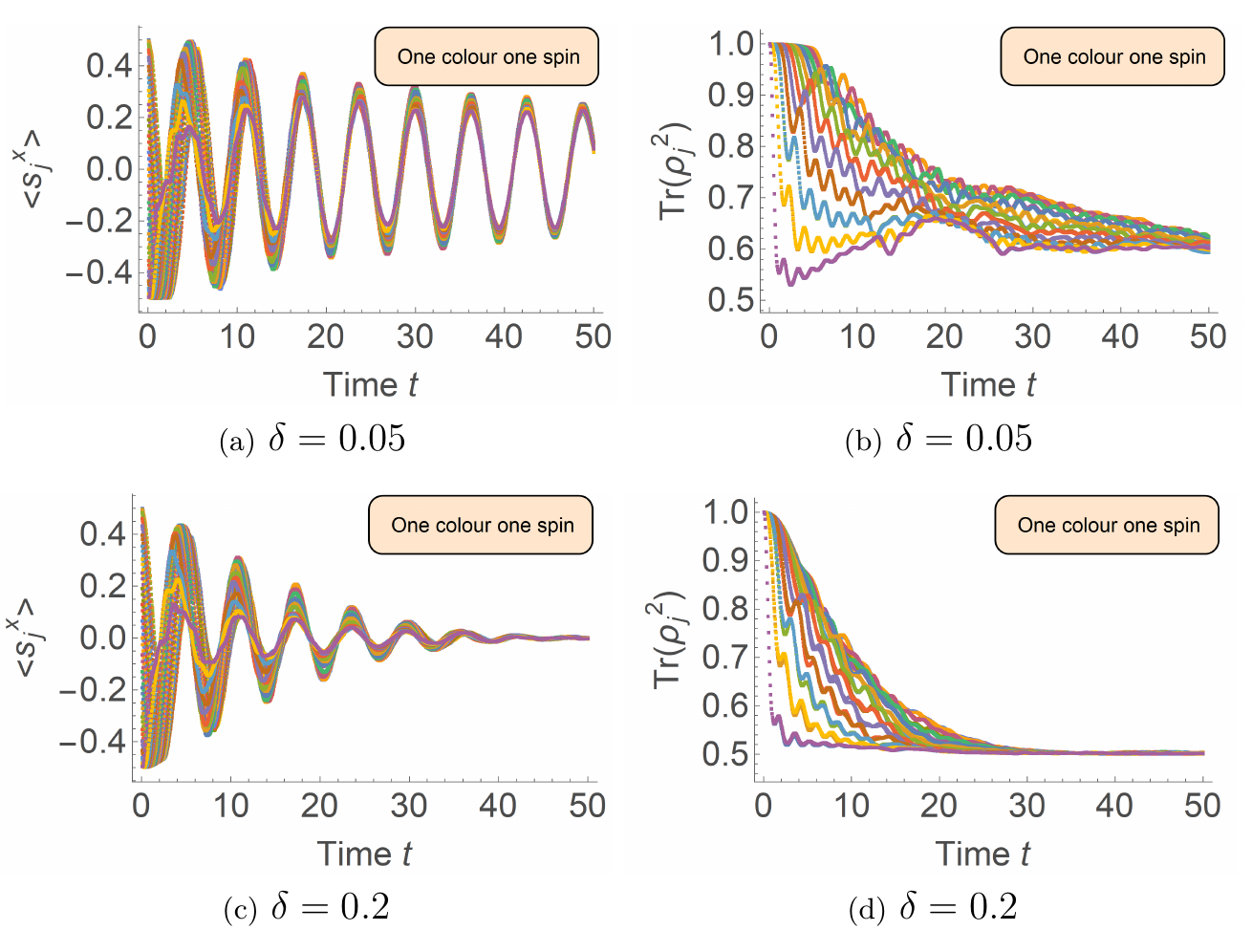}
\caption{\label{fig:9} Time evolutions of initial state $\ket{\psi_B}$ w.r.t.
Hamiltonian \eqref{Hxyz} 
for two values of $\delta$, and $N = 24$, $J = 2$, $h = -1$. 
Videos of \xref{fig:9}(a) and (c) are provided at \cite{VSS:21}.}
\end{figure}

Figure~\xref{fig:9} shows time evolutions for initial state $\ket{\psi_B}$ and two different values of $\delta$.
The magnetization is not a conserved quantity anymore and will therefore decay towards its equilibrium value, 
which is zero in the $xy$-plane for a magnetic field in $z$-direction. 
Our investigations reveal that the larger $\delta$ the faster the spins decay. 
However, we clearly observe that while  decaying the spins still synchronize, see especially \figref{fig:9}(a).
One could say, that the synchronization is a transient phenomenon in such cases since the time scale of synchronization
is shorter than that of the unavoidable decay.

Fugure~\xref{fig:9}(b) and \figref{fig:9}(d) show the purity of the reduced density operators $\op \rho_j$ introduced in 
\eqref{density}.We see that all spins maximally entangle ($\textsl{Tr}   \left( \op \rho_j^2  \right) = 0.5$) which is
equivalent with all individual spin expectation values decay to zero.

Another example of broken symmetry where the spins still synchronize is shown in Appendix \ref{app-2-1} with an
inhomogeneous magnetic field.

\subsection{XX interaction}
\label{sec-4-2}

As comparison we now show a case with XX interaction where the spins do not synchronize. 
The Hamiltonian is defined as

\begin{align}
\op H_{XX} =  &- J \sum_{j = 1}^N  \left( \op s_j^x \op s_{j+1}^x + \op s_j^y \op s_{j+1}^y \right)  - h \sum_{j = 1}^N \op s_j^z \; .
\label{Hxx}
\end{align}

\begin{figure}[h!]
\centering
\includegraphics*[clip,width=1\columnwidth]{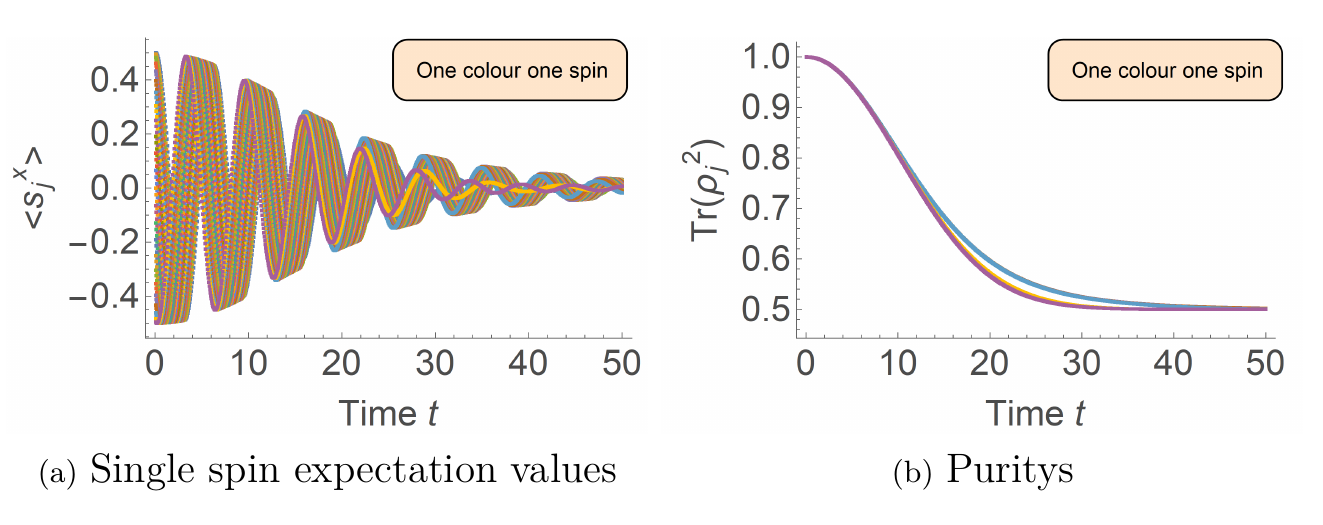}
\caption{\label{fig:10} Time evolution of initial state $\ket{\psi_B}$ w.r.t. 
Hamiltonian \eqref{Hxx} 
with parameters $N = 24$, $J = 0.1$ and $h = -1$. 
The video of \xref{fig:10}(a) can be found at \cite{VSS:21}.}
\end{figure}

Figure~\xref{fig:10} shows a time evolution for initial state $\ket{\psi_B}$. The decay of the transverse magnetization is 
much faster than in the previous subsection, because we are further away from isotropic Heisenberg interactions and the symmetry regarding
the conservation of the transverse magnetization is broken much more strongly,
compare also \cite{JHA:A21}. 
In order to see if 
the spins still synchronize while decaying to zero, 
we choose a much smaller coupling constant $J = 0.1$ instead $J = 2$. 
But we clearly see that the spins do not synchronize while decaying or equivalently the timescale of the decay is much higher than
the timescale of synchronization.

Another example of broken symmetry where the spins do not synchronize is shown in Appendix \ref{app-2-2} with 
dipolar interactions between the spins.

In a future investigation, we plan to study the relation of our findings for the 
translationally invariant XXZ case with the suggestion of a generalized Gibbs ensemble
as the long-time limit of the unitary dynamics induced by dynamical symmetries according to Refs.~\cite{MBJ:PRB20,BTJ:NC19,ZMP:NPB16}.

\section{Summary}
\label{sec-5}

As a conclusion we can say first of all that the conservation of  $M_{\text{trans}}$ 
and $\vec{\op S}^2$ not just slows down the FID, but prevents the free induction from decaying if 
the Hamiltonian only contains isotropic Heisenberg interactions and the Zeeman terms of all
spins are equal (see \figref{fig:2}). This is in accord with Ref.~\cite{UHS:PRB14}.

Furthermore, we demonstrate in detail the interesting phenomenon that the single-spin vector
expectation 
values align in the course of time almost independent of how they are initialized in the $xy$-plane 
(see \figref{fig:3}). It does not matter  if the initial state is a product state of the form \eqref{psi0} or 
if the spins start in an entangled state (see Appendix \ref{app-1-3}). For the process of synchronization
the magnetic field is not necessary, it only induces a (collective) rotation of all spins about the 
field axis (see e.g. \figref{fig:4} and \figref{fig:11}). 
The Heisenberg interactions cause an equilibration process under 
the constraint of conserved quantities. 
We show that after entanglement is maximised (under constraints of conserved quantities) and equilibration is 
completed the spins stay synchronized and fluctuate the less the larger the system is 
(see e.g. \figref{fig:14}, Apppendix \ref{app-1-1}). Moreover, we show that the timescale 
of synchronization is independent of the width $\Delta$ of the variation of 
Heisenberg couplings $J_j$ (see \figref{fig:15}, Apppendix \ref{app-1-1}).

We demonstrate that such a synchronization is not possible with classical 
spins in a closed system (see \figref{fig:7}). 
Moreover we give an example for dissipative 
classical spins which experience a transient synchronization based on ferromagnetism (see \figref{fig:8}). 
We highlight that this is very different 
from the observed quantum mechanical synchronization where the system is closed
and the sign of the Heisenberg interaction does not matter.

In addition, we discuss that the synchronization of spin expectation values is very robust. 
It happens already for small systems ($N = 10$, see \figref{fig:13}, Apppendix \ref{app1})
and for various initial states (see e.g. \figref{fig:6}). 
We find just one exception (initial state $\ket{\psi_C}$) 
which has a special symmetry and in the thermodynamic limit becomes an energy eigenstate (see \figref{fig:5}).
Synchronization also occurs for long-range Heisenberg interactions (see \figref{fig:21}, Appendix \ref{app-2-2}). 
Synchronization is not limited to spins $s = 1/2$. 
In Appendix \ref{app-1-2} we demonstrate that also spins with $s = 1$ synchronize under 
isotropic Heisenberg interactions.

Further on, we provide examples of transient synchronization for systems where symmetries are broken, 
because the time scale of synchronization is shorter than that of equilibration. 
Systems with anisotropic XYZ interactions belong to this set if they are still close 
to the isotropic Heisenberg case (see \figref{fig:9}), or if the symmetry is broken by means of an 
inhomogeneous magnetic field (see \figref{fig:18} and Apppendix \ref{app-2-1}, respectively). 

Finally, we show that spins do not synchronize for interactions that are strongly anisotropic such as 
dipolar interactions (see \figref{fig:20}, Appendix \ref{app-2-2}). 

Our investigations might be helpful for interpreting observations in the context of FID. 
Even if we cannot provide a complete analytical explanation of the phenomenon, 
we think the wide range of numerical results demonstrates that the phenomenon of synchronization
is widespread and robust.

\section*{Acknowledgment}

This work was supported by the Deutsche Forschungsgemeinschaft DFG
355031190 (FOR~2692); 397300368 (SCHN~615/25-1)). 
The authors thank Arzhang Ardavan and Lennart Dabelow 
for fruitful discussions.


%

\appendix

\section{Additions to section \ref{sec-3}}
\label{app1}

\subsection{More data regarding section \ref{sec-3-2}}
\label{app-1-1}

\begin{figure}[h!]
\centering
\includegraphics*[clip,width=1\columnwidth]{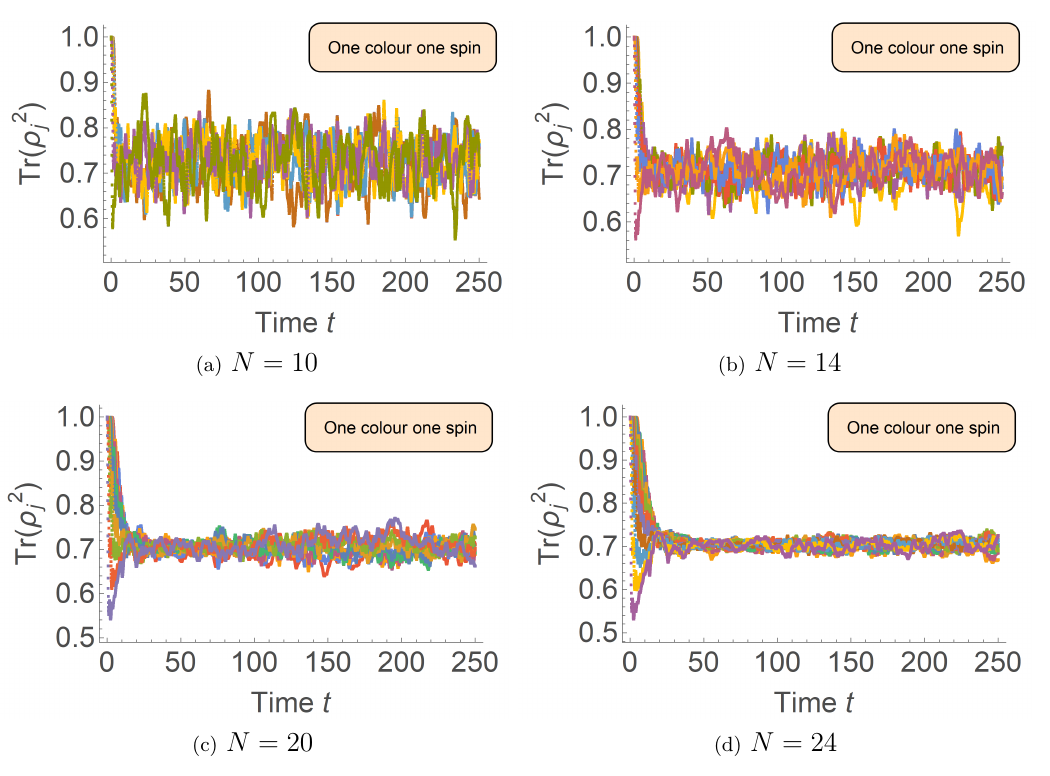}
\caption{\label{fig:11} Purities of the time evolution of initial state $\ket{\psi_B}$ w.r.t.\ Hamiltonian \eqref{H1} 
with isotropic Heisenberg interactions 
and $J_j = 2 \; \forall j$, $h_j \equiv h \; \forall j$. The value of $h$ does not change this figure.}
\end{figure}

Here we present more detailed numerical calculations regarding Section~\ref{sec-3-2}.
In \figref{fig:11} the purity of individual reduced density matrices is shown for different system 
sizes $N$. All Heisenberg couplings $J_j$ are chosen equal and therefore all spins 
entangle in an equal way during equilibration (in contrast to \figref{fig:3}(c)).

It can be clearly seen that the respective purities fluctuate less the 
larger the system is. Thus, in the thermodynamic limit ($N \rightarrow \infty$) 
we expect them to keep the same order of magnitude 
without fluctuating after equilibration.
This figure is completely independent of the strength of the magnetic field $h$, because the Zeeman 
term in Hamiltonian \eqref{H1} does not cause any entanglement between spins.

\begin{figure}[h!]
\centering
\includegraphics*[clip,width=1\columnwidth]{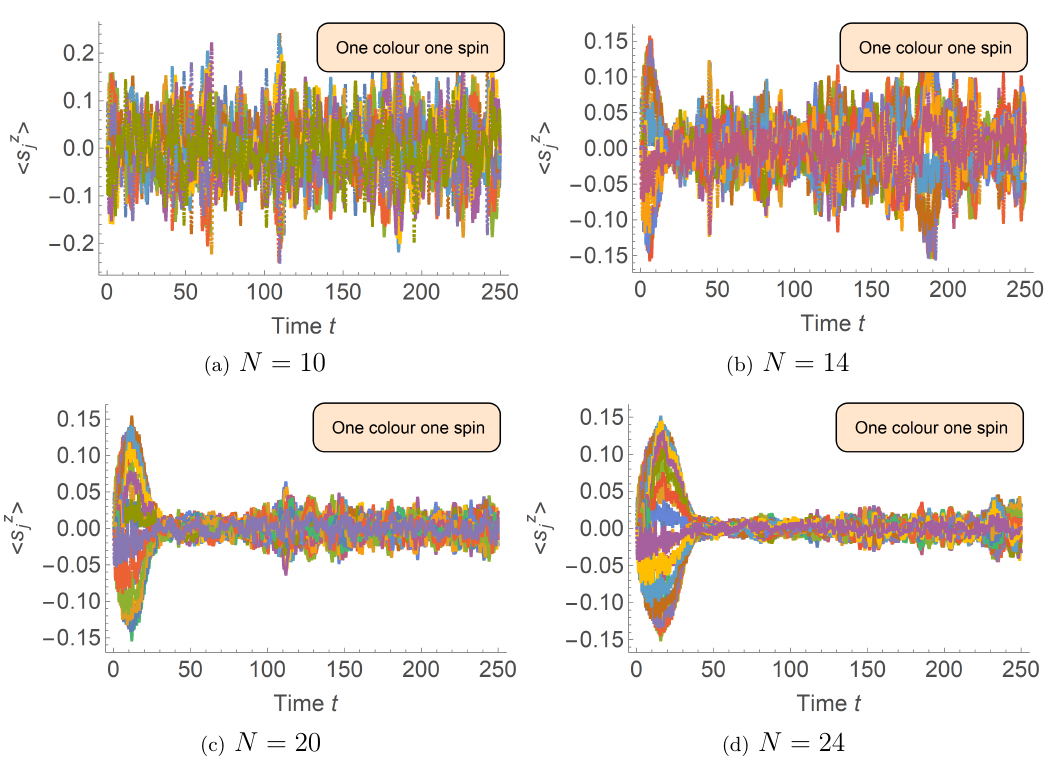}
\caption{\label{fig:12} Single spin expectation values in $z$-direction regarding 
time evolution of initial state $\ket{\psi_B}$ w.r.t.\ Hamiltonian \eqref{H1} 
with isotropic Heisenberg interactions 
and $J_j = 2 \; \forall j$, $h_j \equiv -1 \; \forall j$.}
\end{figure}

\begin{figure}[h!]
\centering
\includegraphics*[clip,width=1\columnwidth]{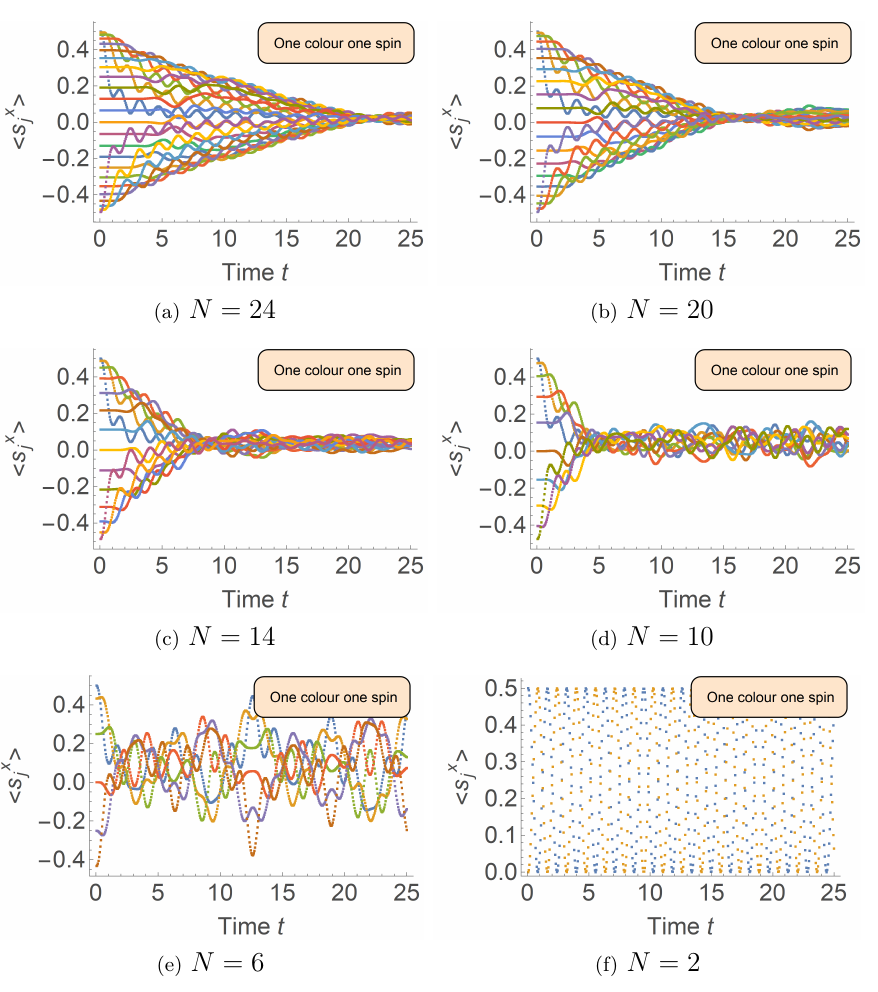}
\caption{\label{fig:13} Single spin expectation values in $x$-direction regarding time evolution 
of initial state $\ket{\psi_B}$ w.r.t.\ Hamiltonian 
\eqref{H1} with isotropic Heisenberg interactions and $J_j  = 2 \;\forall j$, $h_j = 0 \;\forall j$ 
for various system sizes (short time).}
\end{figure}

\begin{figure}[h!]
\centering
\includegraphics*[clip,width=1\columnwidth]{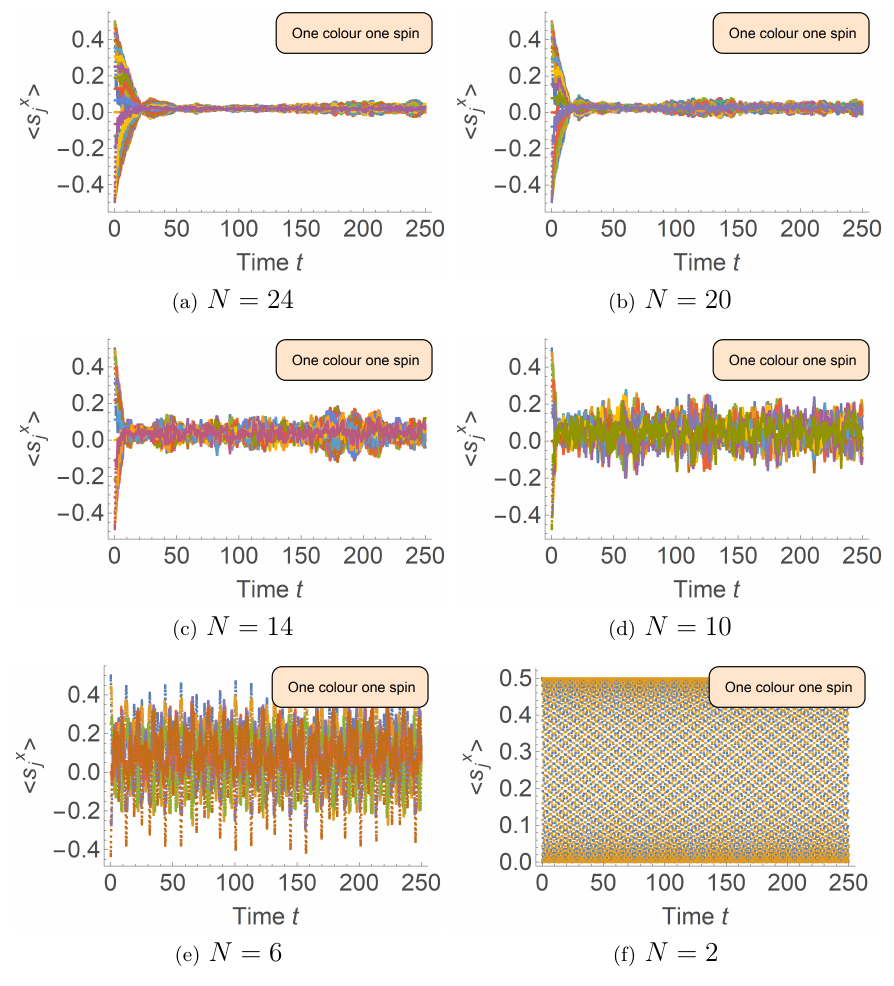}
\caption{\label{fig:14} Same as \figref{fig:13}, but long time.}
\end{figure}

Figure~\xref{fig:12} shows the individual $\langle s_j^z\rangle$ expectation values for exactly the same time evolutions as 
\figref{fig:11}. Initially all these values are zero because the spins are oriented in the $xy$-plane (see \figref{fig:1}). 
During time evolution (especially at the beginning) the 
spins leave the $xy$-plane, but at later times these fluctuations in $z$-direction become small for a larger system
size $N$.

\begin{figure}[h!]
\centering
\includegraphics*[clip,width=1\columnwidth]{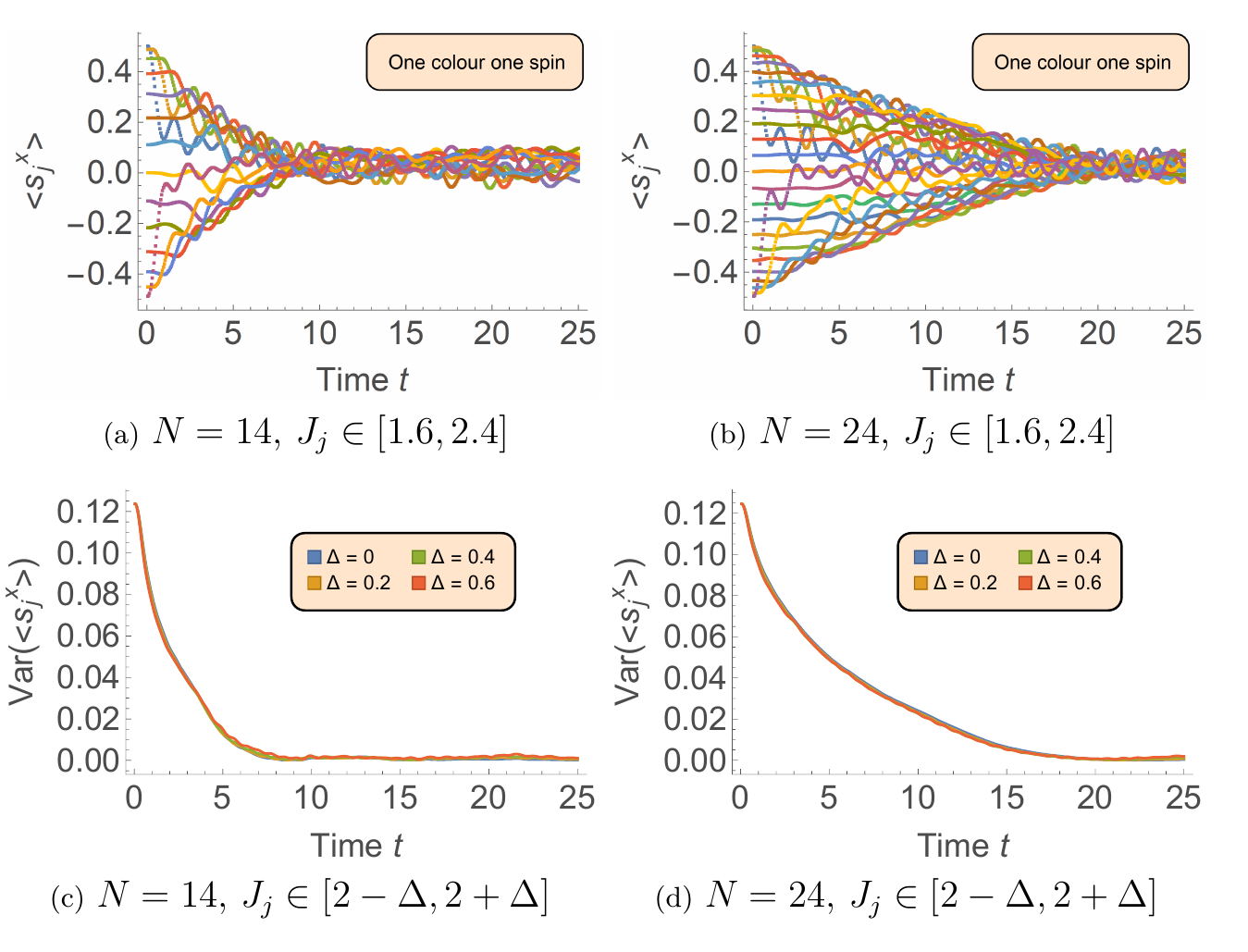}
\caption{\label{fig:15} Time evolution of initial state $\ket{\psi_B}$ w.r.t.\ Hamiltonian 
\eqref{H1} with isotropic Heisenberg interactions from intervals of different width  
$\Delta$, $J_j  \in {[2 - \Delta,2 + \Delta]}$ and different system sizes $N$, $h_j = 0 \;\forall j$.}
\end{figure}

Figure~\xref{fig:13} and \figref{fig:14} show individual spin expectation values for time 
evolutions without magnetic field for different system sizes and for short 
and long time (also for initial state $\ket{\psi_B}$). It can be qualitatively seen that such synchronization of spins can already 
be observed for a small system size of $N = 10$. For $N = 6$ the spins fluctuate 
much and for $N = 2$ the spins permanently point in opposite directions.

We now want to focus on the question how the choice of couplings $J_j$ influences the time 
needed for the spins to align. This is addressed by \figref{fig:15}. 
The couplings $J_j $ are chosen randomly from intervals of different width $\Delta$ and the 
magnetic field is zero. We see that $\Delta$ has a very small impact on 
the time evolution and the behaviour of the spins; the process of alignment and the time it 
takes is very robust. The time to synchronization does only depend on the system 
size $N$ and the mean $J_j$.

\subsection{Spins $\zeta=1$}
\label{app-1-2}

The question arises if the observed synchronization phenomenon also occurs with larger spins than $s = 1/2$. 
To address this question at least partly, we simulate spin rings with spin quantum numbers $\zeta = 1$. 
To simplify our numerical calculation we
form these spins by coupling two spins with $s = 1/2$ each ferromagnetically. We define

\begin{align}
\vec{\op {\zeta}}_j := \vec{\op {s}}_{2j} + \vec{\op {s}}_{2j - 1}
\ .
\label{spin1}
\end{align}

Instead of $N$ spins with $s=1/2$ we have $\tilde{N} = N/2$ spins with $\zeta = 1$. The following Hamiltonian applies 
\begin{align}
\op H =  - \sum_{j = 1}^{\tilde{N}}  J_j \vec{\op {\zeta}}_j \cdot \vec{\op \zeta}_{j+1}  - \sum_{j = 1}^N h_j \op \zeta_j^z \notag \\
- J_f \sum_{j = 1}^{\tilde{N}} \vec{\op {s}}_{2j} \cdot \vec{\op s}_{2j-1}
\ ,
\label{H-modified}
\end{align}

\begin{figure}[h!]
\centering
\includegraphics*[clip,width=1\columnwidth]{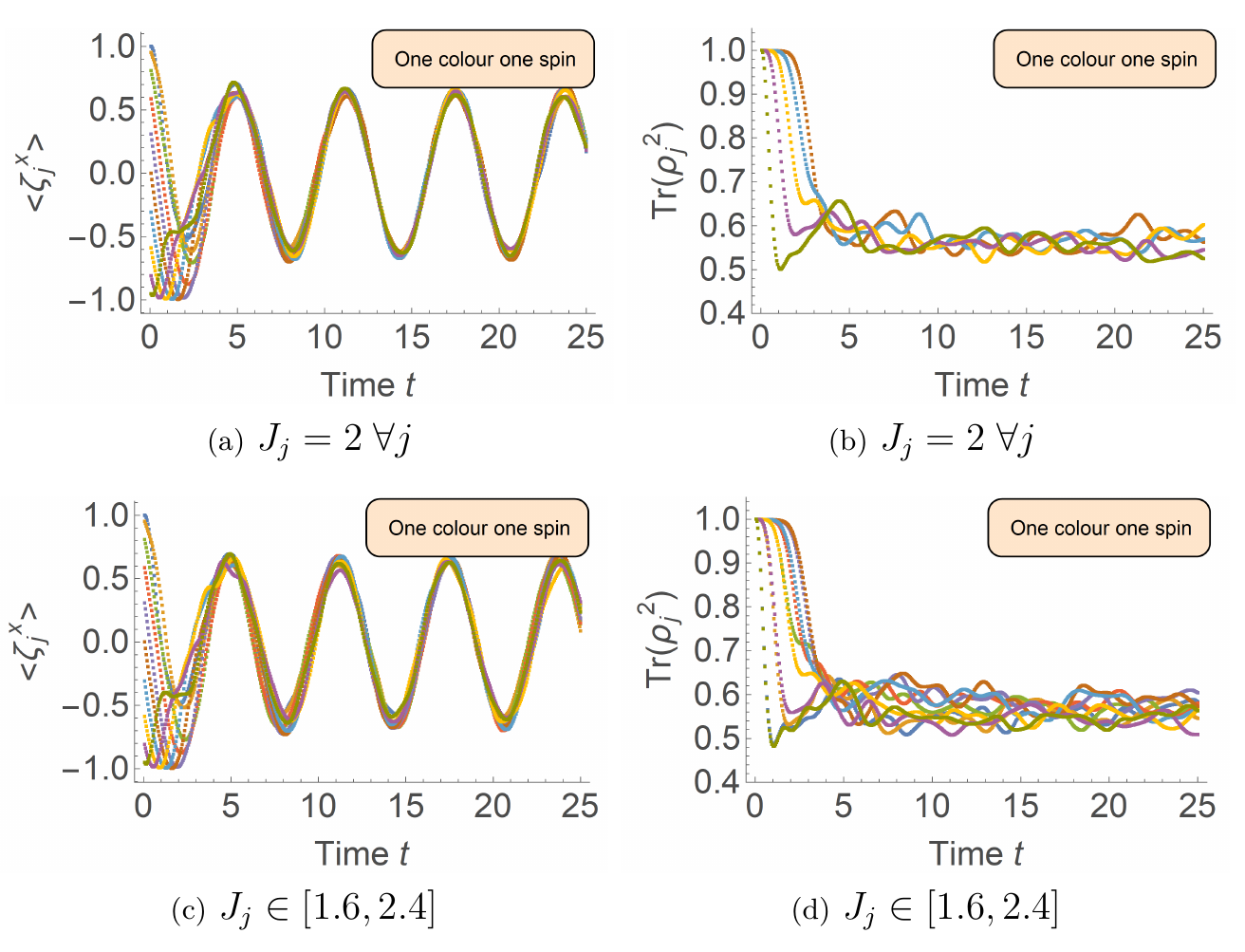}
\caption{\label{fig:16} Time evolution of initial state $\ket{\psi_B}$ w.r.t.\ Hamiltonian 
\eqref{H1} with $h_j = -1 \; \forall j$, $\tilde{N} = 10$ spins with $\zeta =1$ (technically realized as fused pairs of
spins with $s = 1/2$ each).}
\end{figure}

The last term in \eqref{H-modified} is the strong ferromagnetic coupling, $J_f=100\gg |J_j|$, between every two fused spins. 
We initialize 
the state of the system such that those two coupled spins point in the same direction and the $\tilde{N}$ 
spins are fanned out by $180$ degrees. 
The time evolution is shown in \figref{fig:16}, and indeed the spins still synchronize.

We thus speculate that there is no obvious or simple classical limit by which synchronizations would disappear 
when going from small spin quantum numbers to large spin quantum numbers. Also the time scale on which
synchronization happens does not seem to be very different between spins $s=1/2$ and $\zeta = 1$, compare Figs. 
\xref{fig:16}(a) and \xref{fig:13}(d). This question certainly needs further investigations.

\subsection{Starting in an entangled state}
\label{app-1-3}

So far our initial state was always a product state \eqref{psi0}. 
That might give the impression that this could be relevant for the phenomenon of synchronization. 
But this is not the case, as we show with one example in this subsection. 

Consider the state
\begin{align}
\ket{\phi} = \sum_{k = 1}^N c_k \ket{k}
\ ,
\label{initialRandom}
\end{align}
with Gaussian distributed random coefficients and an arbitrary basis $\ket{k}$ of the Hilbert space. 
Such a state will be maximally entangled and all spin expectation values are close to zero. From this we construct our initial state

\begin{figure}[h!]
\centering
\includegraphics*[clip,width=1.0\columnwidth]{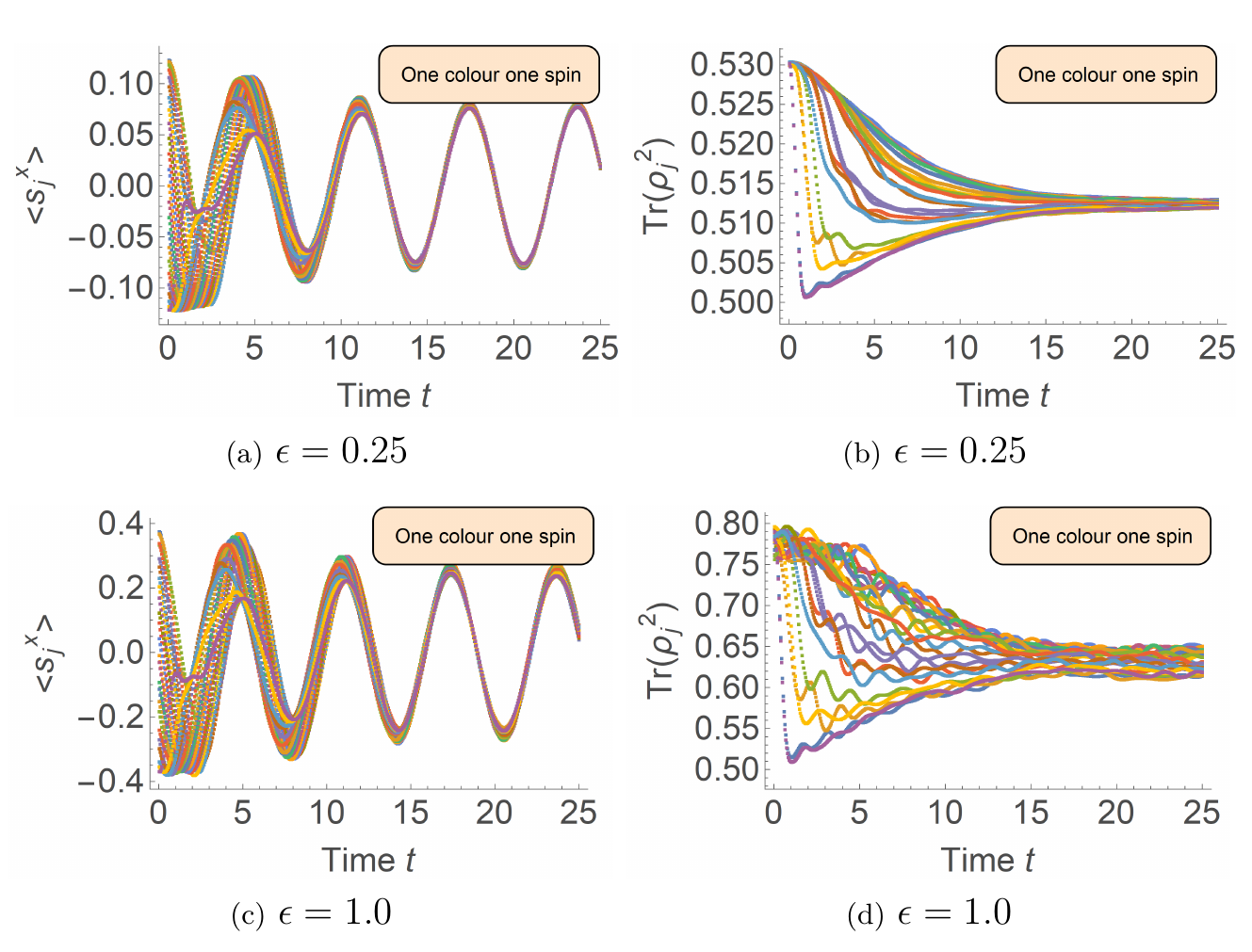}
\caption{\label{fig:17} Time evolution of initial state \eqref{initialEntangled} 
w.r.t.\ Hamiltonian \eqref{H1} with 
isotropic Heisenberg interactions 
and $J_j \in {[1.6,2.4]}$, $h_j = -1 \;\forall j$, $N = 24$ and two different values of $\epsilon$.}
\end{figure}

\begin{align}
\ket{\psi} \propto \left(\prod_{j=1}^N e^{\epsilon \sin{(\alpha_j)} \op s_j^x} e^{\epsilon \sin{(\alpha_j)} \op s_j^y} \right) \ket{\phi}
\ ,
\label{initialEntangled}
\end{align}

with $\alpha_j = \frac{j \pi}{N}$ and $\epsilon$ a parameter for the magnitude of the deflection 
of the spin expectation values 
from zero. This way we create a (still entangled) state with single spin expectations values $\neq 0$ 
in the xy-plane fanned out by $180$ degrees.
\figref{fig:17} shows associated time evolutions and that the spins do indeed still synchronize, 
starting in an entangled state \eqref{initialEntangled}.

\section{Additions to section \ref{sec-4}}
\label{app2}

In addition to sec.~\xref{sec-4} we present more cases of how the spins behave with broken symmetry 
(without the conserved quantities 
$M_{\text{trans}}$ and $\vec{\op S^2}$).

\subsection{Inhomogeneous magnetic field}
\label{app-2-1}

As shown in \eqref{eqkommu2} all magnetic fields $h_j$ have to be equal or the conserved quantities are broken. 

\begin{figure}[h!]
\centering
\includegraphics*[clip,width=1\columnwidth]{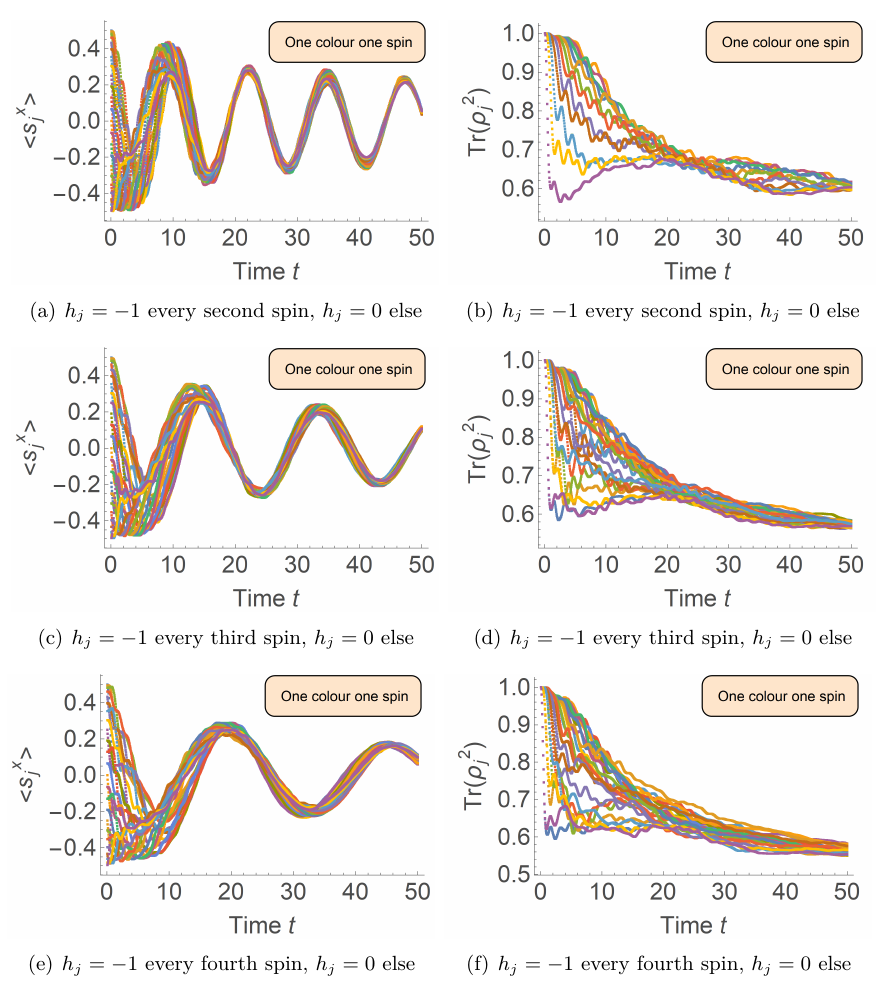}
\caption{\label{fig:18}  Time evolution of initial state $\ket{\psi_B}$ w.r.t.\ Hamiltonian 
\eqref{H1} with isotropic Heisenberg interactions and $J_j  = 2 \;\forall j$, $N = 24$ 
and for different configurations of $h_j$. 
Videos of \xref{fig:18}(a) and (e) are provided at \cite{VSS:21}.}
\end{figure}

Figure~\xref{fig:18} shows time evolutions of initial state $\ket{\psi_B}$ where only a few spins see a magnetic 
field $h_j = h  \neq 0$ and all others fields are zero. Surprisingly the spins do still synchronize an precess together while they decay. 
The spins without magnetic field are carried with the others. 

The frequency $\tilde{h}$ of the collective precession decreases the more
spins there are with $h_j = 0$. This can be viewed as every spin sees a mean field $\tilde{h} = \sum_j h_j/N$. 
From \figref{fig:18}(a) to \figref{fig:18}(e) 
the number of magnetic fields $h_j = h$ is halved and the precession frequency $\tilde{h}$ also halves.

\begin{figure}[h!]
\centering
\includegraphics*[clip,width=1\columnwidth]{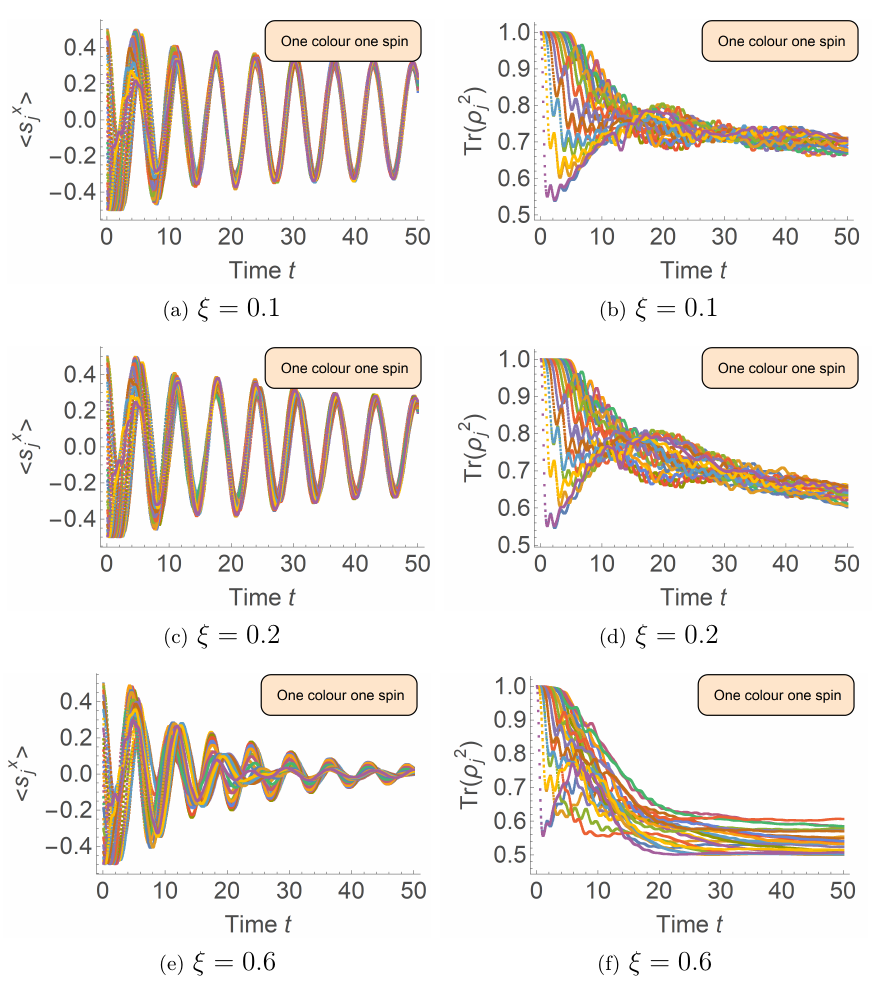}
\caption{\label{fig:19}  Time evolution of initial state $\ket{\psi_B}$ w.r.t.\ Hamiltonian 
\eqref{H1} with isotropic Heisenberg interactions, $J_j  = 2 \;\forall j$, $N = 24$ 
and $h_j \in {[-1 - \xi, -1 + \xi]}$.
Videos of \xref{fig:19}(a) and (e) are provided at \cite{VSS:21}.}
\end{figure}

Another way of breaking the symmetry is by choosing random magnetic fields. Figure~\xref{fig:19} shows time 
evolutions for initial state $\ket{\psi_B}$ where the magnetic fields $h_j$ are drawn at random from a
distribution of different width $\xi$. The spins do still synchronize up to large values of $\xi$ up to the point where the decay 
of the transverse magnetization is faster than the synchronization. 

\subsection{Dipolar and long range interactions}
\label{app-2-2}

We now investigate how initial state $\ket{\psi_B}$ behaves when all spins (not only neighbors) 
interact with dipolar interactions or just long-range Heisenberg interactions. 

The Hamiltonian of a dipolar interacting spin system is given by

\begin{align}
\op H  = & \sum_{j = 1}^{N}  \sum_{k = j+1}^{N} \frac{\lambda}{r_{jk}^3} \left( \op{\vec{s}}_{j} \cdot \op{\vec{s}}_{k} - \frac{3 ( \op{\vec{s}}_{j} \cdot \vec{r}_{jk}  ) ( \op{\vec{s}}_{k} \cdot \vec{r}_{jk})}{r_{jk}^2} \right) \notag \\ 
&- h \sum_{j = 1}^N  \op s_j^z \;.
\label{eqHdipolar}
\end{align}

We use the parameter $\lambda$ to tune the strength of the interaction.
We also need spatial coordinates $\vec{r}_j$ for all spins. 
For simplicity we arrange them on the unit circle

\begin{align}
 \vec{r}_j= \left(\begin{array}{c} \cos(2 \pi j/N) \\ \sin(2 \pi j/N) \\ 0 \end{array}\right) \; .
\label{eqcoordinates}
\end{align}

\begin{figure}[ht!]
\centering
\includegraphics*[clip,width=1\columnwidth]{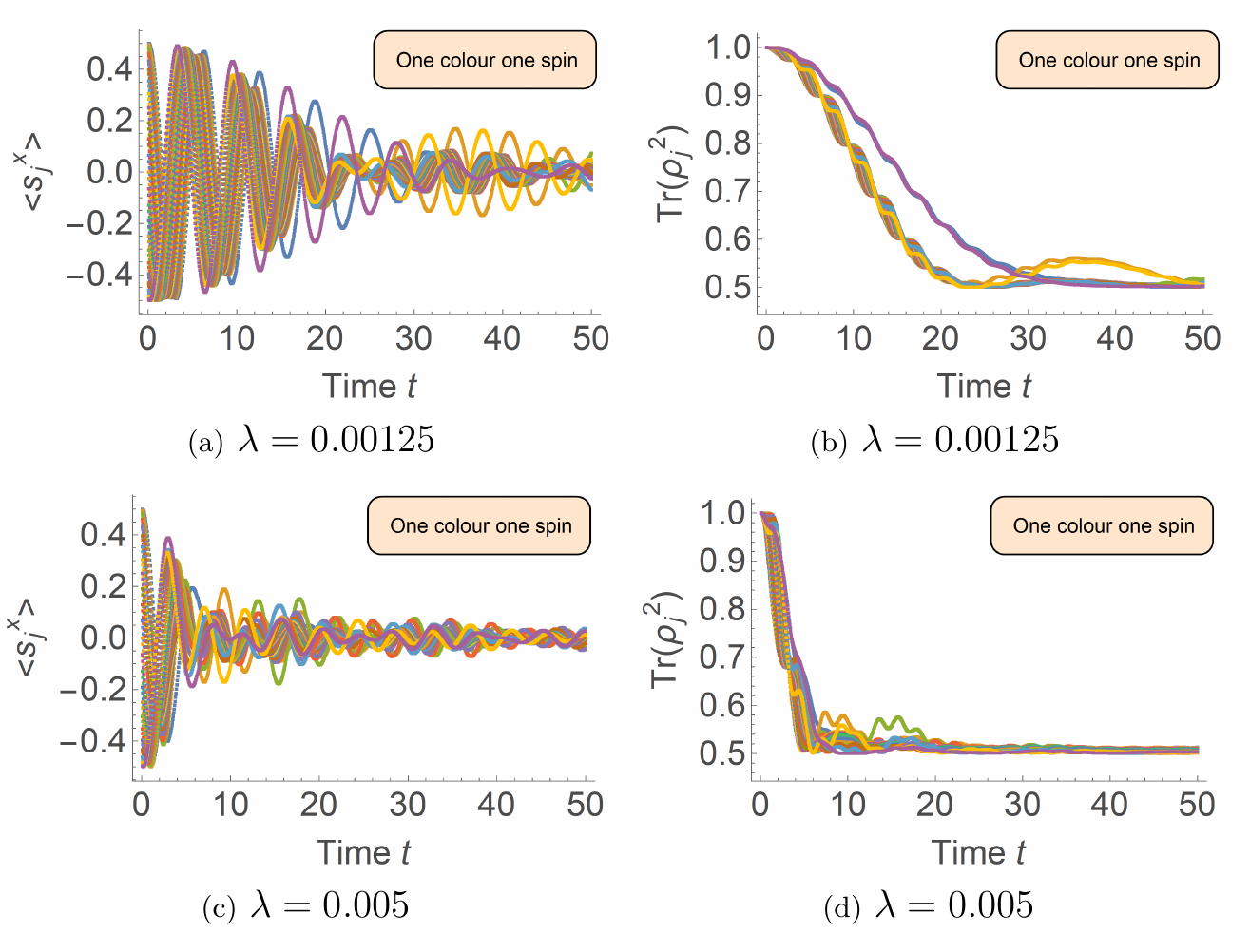}
\caption{\label{fig:20}  Time evolution of initial state $\ket{\psi_B}$ w.r.t.\ Hamiltonian 
\eqref{eqHdipolar} with coordinates \eqref{eqcoordinates}, $N = 24$, and for different parameters $\lambda$.
The video of \xref{fig:20}(a) can be found at \cite{VSS:21}.}
\end{figure}

Figure~\xref{fig:20} shows time evolutions for initial state $\ket{\psi_B}$ for two different parameters $\lambda$.
As expected the spin expectation values decay the faster the larger the parameter $\lambda$ is. But even for a small
$\lambda$ in \figref{fig:20}(a) the spins do not synchronize at all. 
This can also be seen in the related video at \cite{VSS:21}. 
Dipolar interactions are highly anisotropic, therefore, the conservation of the transverse magnetization is broken strongly.

\begin{figure}[ht!]
\centering
\includegraphics*[clip,width=1\columnwidth]{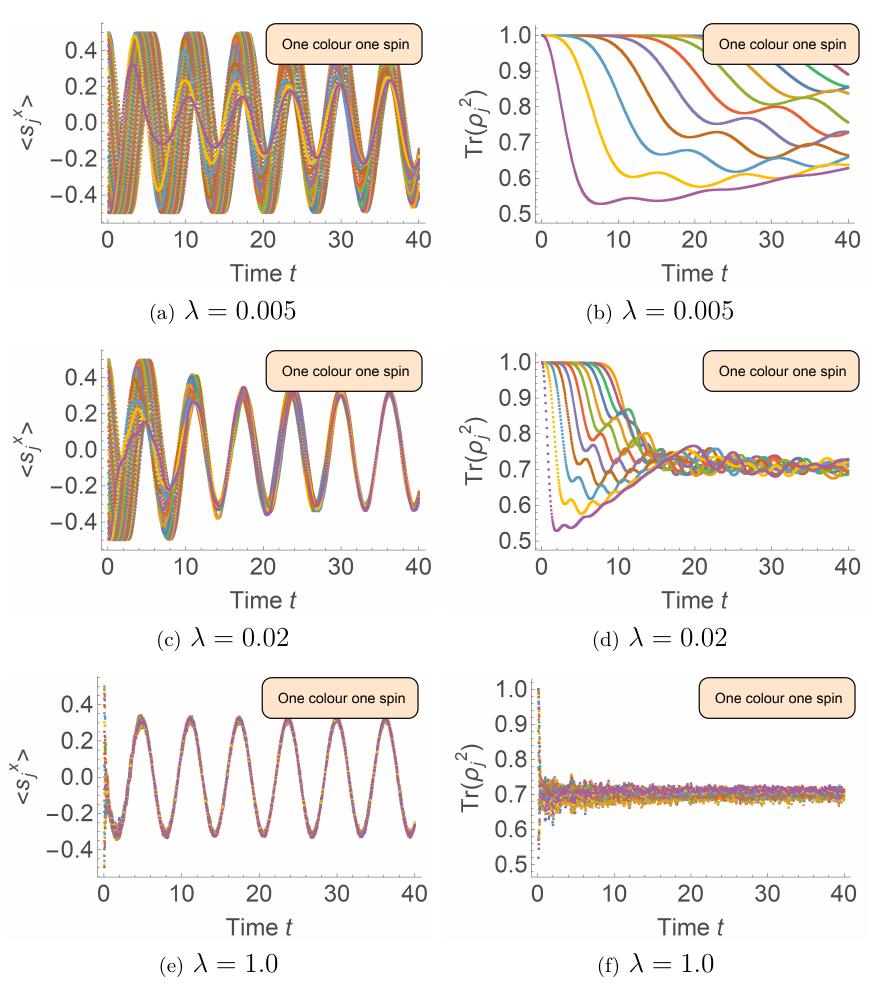}
\caption{\label{fig:21}  Time evolution of initial state $\ket{\psi_B}$ w.r.t.\ Hamiltonian 
\eqref{eqHeiLong} with coordinates \eqref{eqcoordinates}, $N = 24$, and for different parameters $\lambda$.}
\end{figure}

Finally, we want to test if synchronization under isotropic Heisenberg couplings is limited to nearest-neighbor interactions 
or holds under long-range interactions. To this end, we take as an example the above Hamiltonian \eqref{eqHdipolar} 
and remove the anisotropic parts

\begin{align}
\op H  =  \sum_{j = 1}^{N}  \sum_{k = j+1}^{N} \frac{\lambda}{r_{jk}^3} \op{\vec{s}}_{j} 
\cdot \op{\vec{s}}_{k} - h \sum_{j = 1}^N  \op s_j^z \;.
\label{eqHeiLong}
\end{align}

Again we tune the interaction strength by the parameter $\lambda$. \figref{fig:21} shows calculations for very small to larger values of $\lambda$. For $\lambda = 0.005$ in \figref{fig:21} the time scale shown is not sufficient for the spins to synchronize, whereas in \figref{fig:20}(c) 
there  is enough time  for the magnetization to decay completely. This shows that the magnetization decay through 
the anisotropic terms of the dipolar interaction happens on a much faster time scale than the isotropic part causes synchronization. 
Choosing $\lambda$ significantly larger in \figref{fig:21} leads to synchronization in the given time frame. 

One finding of this chapter is that the synchronization effect is more general, and it does not only appear for 
nearest-neighbor isotropic Heisenberg interactions.

\end{document}